\documentclass[aps,prd,twocolumn,superscriptaddress,nofootinbib,longbibliography]{revtex4-2}

\usepackage{amsmath, amssymb}
\usepackage{siunitx}
\usepackage{graphicx}
\usepackage{tabularx}
\usepackage{array}
\usepackage{hyperref}
\usepackage{url}
\usepackage{float}
\usepackage{booktabs}
\usepackage{orcidlink}
\usepackage[normalem]{ulem}
\providecommand{\pasp}{Publ. Astron. Soc. Pac.}
\hypersetup{colorlinks=true, allcolors=blue}
\usepackage{subcaption}
\usepackage{ulem}

\begin{document}

\title{Untangling Extreme Mass Ratio Inspirals and Massive Black Hole Binaries in LISA Data}
\author{Rishav Agrawal\,\orcidlink{0009-0003-8007-5802}}
\email{rishavrste@u.nus.edu}
\affiliation{Department of Physics, National University of Singapore, 21 Lower Kent Ridge Rd, Singapore 119077}

\author{Cl\'emence Allietta\,\orcidlink{0009-0008-0063-0420}}
\affiliation{CNES, Centre Spatial de Toulouse, 18 avenue Édouard Belin, 31401 Toulouse Cedex 9, France}
\affiliation{Laboratoire des 2 Infinis – Toulouse (L2IT-IN2P3),
Universitè de Toulouse, CNRS, UPS, F-31062 Toulouse Cedex 9, France}
\author{Ollie Burke,\orcidlink{0000-0003-2393-209X}}

\affiliation{Laboratoire des 2 Infinis – Toulouse (L2IT-IN2P3),
Universitè de Toulouse, CNRS, UPS, F-31062 Toulouse Cedex 9, France}
\affiliation{School of Physics and Astronomy, University of Glasgow, Glasgow G12 8QQ, United Kingdom}
\author{Alvin J. K. Chua\,\orcidlink{0000-0001-5242-8269}}
\affiliation{Department of Physics, National University of Singapore, 21 Lower Kent Ridge Rd, Singapore 119077}
\affiliation{Department of Mathematics, National University of Singapore, 21 Lower Kent Ridge Rd, Singapore 119077}

\begin{abstract}

The Laser Interferometer Space Antenna (LISA) will observe sources in the millihertz regime, such as Galactic Binaries (GBs), Massive Black Hole Binaries (MBHBs), and Extreme Mass Ratio Inspirals (EMRIs), so its data stream will be a superposition of an unknown number of gravitational-wave sources. We consider the joint inference of various representative combinations of MBHB and EMRI signals in LISA data (with assumed instrumental noise and stochastic GB foreground). Our Bayesian analyses confirm the expectation of weak inter-source coupling between EMRIs, and between EMRIs and MBHBs, despite strong intra-source parameter correlations. Weak EMRI–MBHB coupling enables a hierarchically sequential approach to EMRI and MBHB inference, rather than fully joint inference. We further show that imperfect MBHB subtraction, arising from missing modes or overlapping MBHBs, has only a weak impact on EMRI inference. For multiple MBHBs, we mitigate inference errors through a joint MBHB fit. We demonstrate this strategy on simulated LISA data containing noise, three MBHBs, and three EMRIs, jointly fitting and subtracting the MBHBs before sequentially inferring the EMRIs --- ultimately observing little to no bias on all recovered astrophysical parameters. The notions described throughout our work are highly relevant to informing optimal strategies for the global-fit-based pipelines that are currently being developed for LISA.

\end{abstract}

\maketitle

\tableofcontents

\section{Introduction}

Ground-based gravitational-wave (GW) detectors, namely the LIGO-Virgo-KAGRA (LVK) network, have found great success in detecting transient sources from compact binary systems \cite{KAGRA:2021vkt,LIGOScientific:2025slb, LIGOScientific:2026wfs}. Building on that success, the Laser Interferometer Space Antenna (LISA) is an approved ESA--NASA GW mission comprising three near-identical spacecraft in a triangular configuration, with each arm measuring approximately 2.5 million kilometers, giving the means to probe the millihertz GW regime spanning roughly from 0.1 mHz to 1 Hz.

Unlike the LVK network, where the majority of sources are transient, LISA will observe a superposition of signals from a non-exhaustive list of GW sources, including Extreme Mass Ratio Inspirals (EMRIs), Massive Black Hole Binaries (MBHBs), and Galactic Binaries (GBs) \cite{LISA:2024hlh}. EMRIs and MBHBs are some of the most important sources to characterize. An EMRI consists of a stellar-mass compact $m_1 \sim 10^{(0 - 2)}M_{\odot}$ object inspiraling around a supermassive black hole (SMBH) with $M \sim 10^{(5 - 7)}M_{\odot}$. In this system, the Signal-to-Noise Ratio (SNR) slowly accumulates as the inspiral phase emits GWs in the detectable region, taking around $10^4 - 10^6$ orbits before the plunge \cite{Hughes:2000ssa,Hinderer:2008dm,Hughes:2016xwf,Barack:2018yvs,Pound:2021qin}. An MBHB system consists of a two massive black holes $(m_1,m_2)\sim 10^{(4 - 7)} M_{\odot}$ with more comparable mass-ratios $q = m_2/m_1 \sim 1$--$100$, observable on the time-scales of days to months. Unlike EMRIs, whose SNR accumulates slowly throughout the observation period, the majority of the SNR is concentrated during the merger phase \cite{Colpi:2019yzd,LISA:2024hlh,Marsat:2020rtl}.

LISA observations of MBHBs can elucidate their evolution history and help us understand their structure formation \cite{Sesana:2010wy, Volonteri:2021sfo}. Given the high-precision nature of GW astronomy, these sources can be used as standard sirens, providing precise measurement of distance and the Hubble constant \cite{Schutz:1986gp, Holz:2005df}. Similarly, EMRIs constitute an independent class of GW standard sirens, including in strongly-lensed configurations \cite{Toscani:2023gdf, Laghi:2021icm}. EMRI-based cosmology can be combined with MBHB analysis to break cosmological parameter degeneracies \cite{Laghi:2026ujb}. Moreover, these sources will also allow us to test the validity of General Relativity in the low-frequency regime \cite{Gair:2012nm}. Some recent works have used EMRIs to probe for massless scalar charges in modified-GR theories \cite{ Speri:2024qak, Barsanti:2026ulr}. Realizing these scientific objectives, however, hinges on accurate parameter estimation of the source parameters, as unmodeled physical effects and waveform systematics~\cite{Burke:2023lno, Khalvati:2025znb} can bias tests of General Relativity~\cite{Speri:2021psr,Garg:2024qxq}.

Although MBHB and EMRI signals have very high information content, enabling precise parameter estimation, the task is generally challenging due to the superposition of signals from all sources \cite{LISA:2024hlh}. Although detection rates are uncertain, we could expect up to 10,000 compact binaries, several hundred EMRIs, and of order a hundred MBHBs in the LISA detectable frequency band. The inverse problem of inferring the parameters of each source from this superposed signal, without bias, is highly nontrivial. Several efforts are underway to develop iterations of a \textit{global-fit} pipeline that simultaneously detects and characterizes all GW sources, along with the noise \cite{Littenberg:2020bxy,Littenberg:2023xpl,Katz:2024oqg, Strub:2024kbe,Deng:2025wgk}.

Existing LISA global-fit pipelines have been tested on the LISA Data Challenge 2A (\textit{Sangria}) data set, which contains MBHBs, GBs (predominantly white-dwarf (WD) binaries), and instrumental noise, but no EMRIs. Since GB signals are quasi-monochromatic, their signal can be divided into narrow frequency segments analyzed using reversible-jump MCMC~\cite{Littenberg:2020bxy}. The \texttt{GLASS}~\citep{Littenberg:2023xpl} and \texttt{Erebor}~\citep{Katz:2024oqg} pipelines, the latter utilizing GPU acceleration for improved computational efficiency, update MBHB, GB, and noise parameters in separate blocks. Meanwhile, \citet{Deng:2025wgk} combine a maximum-likelihood search with Gibbs-like block MCMC sampling, whereas \citet{Strub:2024kbe} employ a sequential maximum-likelihood approach. Existing global-fit approaches generally exploit the expectation of weak correlations between different source classes, an assumption that has not yet been tested when EMRIs are present in the data.

A fraction of EMRIs and GBs in the LISA band will remain unresolved and contribute astrophysical confusion foregrounds to the LISA data. Previous studies have shown that the unresolved EMRI background can affect the LISA sensitivity \citep{Barack:2005aj,Bonetti:2020jku}, while recent analyses have demonstrated departures from Gaussianity and stationarity in the GB foreground in the 2--10 mHz band \citep{Buscicchio:2024wwm}, as well as in the EMRI background, with the magnitude of these deviations depending on the brightness of the background \citep{Piarulli:2024yhj}.

As for resolvable sources, \citet{Chua:2021aah} conjectured that EMRIs are unlikely to self-confuse: pairwise overlaps among 200 detectable EMRIs were strongly peaked around zero, while the root-mean-square overlap grows slowly and approximately linearly with the number of EMRIs, suggesting that self-confusion remains negligible for $N\lesssim10^4$. In contrast, \citet{Khukhlaev:2025xiz} showed that bright, resolvable WD binaries can interact with EMRIs, producing false detections and biasing EMRI parameter inference. This motivates a hierarchical treatment in which bright WD binaries are characterized and subtracted before performing EMRI inference. Although relevant for the LISA mission, we neglect the impact of EMRI and GB confusion in this work to focus on inference from resolvable MBHBs and EMRIs. To our knowledge, parameter-estimation studies have not directly investigated the cross-talk between MBHBs and EMRIs. This study aims to address this gap and inform the development of inference strategies for incorporating MBHBs and EMRIs within future global-fit pipelines.

\subsection{Outline and Main Findings}
In this work, we investigate interactions between EMRIs and MBHBs using a full Bayesian analysis across various representative signal combinations. This study serves as a step towards the development of a global-fit pipeline for LISA. Specifically, we ask how the two signal types interact, how these interactions affect the recovery of the true source parameters, and which inference strategies can mitigate these effects while minimizing computational cost. Since full Bayesian inference for even a single EMRI is computationally expensive, we do not attempt a population-level injection campaign. Instead, we select representative configurations designed to probe distinct regimes of source interaction.

We begin in Section~\ref{just_EMRI} by considering EMRI signals in isolation (without MBHBs, GBs and instrumental noise). This allows us to investigate self-confusion among EMRI signals and to characterize the correlations that arise between multiple EMRI sources. Subsequently, in Section~\ref{EMRI_MBH}, we consider a two-source scenario consisting of one EMRI and one MBHB. This analysis aims to elucidate the interactions between these two signal classes and to determine how the order and manner in which the signals are modeled affect the recovery of their source parameters. 

Next, in Section~\ref{2MBH_1EMRI}, we extend our analysis to a three-source configuration comprising two MBHBs and one EMRI. In this setting, the two MBHB mergers may occur sufficiently close together in time that their signals become difficult to resolve, thereby complicating parameter inference. We investigate strategies for mitigating the resulting inference errors. Section~\ref{missing mode} then examines the systematic error introduced by omitting waveform modes in MBHB inference and assesses how these errors propagate into the subsequent inference of EMRI parameters. Finally, in Section~\ref{lisa_data_analysis}, we consider a more realistic LISA data stream comprising three MBHBs, three EMRIs, and instrumental noise sampled from a PSD that includes a confusion foreground from unresolved WD binaries. We use this data stream to test the efficacy and robustness of the inference methods used throughout the paper.

For readers seeking a concise overview, our main findings are summarized below:
\begin{enumerate}
\item \emph{Inter-source coupling is weak for the configurations considered.} We find negligible inter-source parameter correlations among EMRIs and between EMRIs and MBHBs, supporting sequential EMRI inference as a computationally efficient alternative to fully joint inference.
\item \emph{MBHBs should generally be characterized first.} An unmodeled MBHB can produce significant deviations in EMRI parameter recovery, whereas the presence of an EMRI has a much smaller impact on MBHB inference. This ordering remains effective even in challenging configurations involving a loud EMRI or a nearly coincident EMRI plunge and MBHB merger.

\item \emph{Temporally overlapping MBHB mergers benefit from joint inference.} When two MBHB mergers occur close together in time, sequential inference can introduce noticeable parameter deviations, while jointly fitting the MBHBs substantially improves parameter recovery.

\item \emph{Large deviations in MBHB parameters do not necessarily propagate to the EMRI.} Missing higher-order MBHB modes can substantially shift the recovered MBHB parameters, while producing considerably smaller effects on the subsequent EMRI inference. The downstream impact is primarily determined by the residual waveform remaining after MBHB subtraction.
\end{enumerate}

\section{Waveform models}
In this section, we summarize the waveform models used to generate the EMRI and MBHB signals. 
\subsection{EMRIs}
For an EMRI on an eccentric equatorial orbit, the source-frame waveform can be expressed in terms of the complex, dimensionless time-domain strain $
h(t) = h_{+} - i h_{\times}$, 
where \(h_{+}\) and \(h_{\times}\) are the two transverse-traceless GW polarizations. The strain can be decomposed into a sum over harmonic modes as
\[h = \frac{\mu}{d_L}
\sum_{lmn}A_{lmn}(t)
S_{lmn}(t,\theta)
e^{im\phi}
e^{-i\Phi_{mn}(t)}
\]
where \(t\) is the arrival time of the GW at the Solar System barycenter (SSB), \(\theta\) is the source-frame polar angle, \(\phi\) is the source-frame azimuthal angle, \(d_L\) denotes luminosity distance, and \(\{l,m,n\}\) are frequency harmonic describing the mode decomposition in the frequency domain of the signal. The indices \(l\), \(m\), and \(n\) stand for the orbital angular momentum, azimuthal, and
radial modes, respectively. For eccentric equatorial orbits, the phase of each harmonic takes the form of $\Phi_{mn}
= m\Phi_{\phi} + n\Phi_{r}.$

Next, the $A_{lmn}$ and $S_{lmn}$ are Teukolsky mode amplitude and spin-weighted spheroidal harmonics. These are solved using the Teukolsky formalism in Kerr Spacetime. Further details about the derivation can be found in \cite{Hughes:2021exa}. Since we restrict our analysis to equatorial orbits, the polar motion is absent, and hence no independent polar phase $\Phi_{\theta}$ or associated $k$-harmonics are required, thus the initial polar phase has no effect on the waveform. 

In this analysis, we used the \texttt{FastKerrEccentricEquatorialFlux} model provided in the \texttt{FastEMRIWaveform} (FEW) Python package~\cite{Chua:2020stf,Katz:2021yft,Chapman-Bird:2025xtd}, which supports GPU acceleration and computes gravitational waveforms for EMRIs in eccentric equatorial inspirals around a Kerr black hole. GPU acceleration reduces the waveform generation time to around 100 milliseconds, making parameter sampling computationally feasible.

The EMRI waveform $h$ in the \texttt{KerrEccentricEquatorial} model depends on the 12 EMRI parameters as described in the Table~\ref{tabel:table_emri_param}. As we only treat equatorial orbits, the inclination of the secondary body does not evolve and is kept fixed~\cite{Chapman-Bird:2025xtd}. The first five parameters, \{$M$, $\mu$, $a$, $p_0$, $e_0$\}, together with the initial phase angles \{$\Phi_{\phi_0}$, $\Phi_{r_0}$\}, can be considered intrinsic parameters, while the remaining ones are extrinsic parameters. In this analysis, we infer 12 EMRI parameters: \{$M$, $\mu$, $a$, $p_0$, $e_0$, $d_L$, $q_S$, $\phi_S$, $q_K$, $\phi_K$, $\Phi_{\phi_0}$, and $\Phi_{r_0}$\}.
\begin{table}[ht]
\centering
\caption{Definition of the parameters describing an EMRI}
\label{tab:emri_parameters}
\begin{tabular}{cl}
\toprule
\textbf{Parameter} & \textbf{Description} \\
\midrule

$M$ & Mass of the central SMBH \\

$\mu$ & Mass of the secondary BH \\

$a$ & Dimensionless spin of the central black hole \\

$p_0$ & Initial semi-latus rectum \\

$e_0$ & Initial eccentricity \\

$d_L$ & Luminosity distance to source \\

$q_S$ & Sky location polar angle \\

$\phi_S$ & Sky location azimuthal angle \\

$q_K$ & Initial BH spin polar angle \\

$\phi_K$ & Initial BH spin azimuthal angle \\

$\Phi_{\phi_0}$ & Initial azimuthal fundamental phase \\

$\Phi_{r_0}$ & Initial radial fundamental phase \\
\bottomrule
\end{tabular}
\label{tabel:table_emri_param}
\end{table}
\subsection{MBHBs}
 \label{MBHB_waveform_des}
The waveform of an MBHB is composed of multiple modes. In the frequency domain, it is of the form :
\begin{equation}
h(f) = \sum_{lm} A_{lm}(f)e^{i\Phi_{lm}(f)},
    \label{eq:mbhb_waveform}
\end{equation}
where $A_{lm}$ and $\Phi_{lm}$ are respectively the amplitude and
phase of mode $(l,m)$.  For this project, we used the \texttt{PhenomHM} \cite{London:2017bcn} model provided by the \texttt{BBHx} package which leverages GPU acceleration~\cite{Katz:2020hku}.
\texttt{PhenomHM} is an inspiral-merger-ringdown gravitational wave model for spinning binary black holes utilizing the Stationary Phase Approximation (SPA) in its construction, which generates the waveform in the frequency domain. The idea behind the phenomenological approach is to create a finite waveform family based on available Numerical Relativity (NR) data and to deduce an analytical model from this limited number of waveforms \cite{Ajith:2007qp, Santamaria:2010yb,Ajith:2009bn,Hannam:2013oca, Husa:2015iqa,Khan:2015jqa}.

The \texttt{BBHx} package allows faster waveform evaluation with six available $(l,m)$ modes: $(2,2)$, $(3,3)$, $(4,4)$, $(2,1)$, $(3,2)$, and $(4,3)$, with the $(2,2)$ being dominant. The relative contribution of the higher-order modes depends on the properties and orientation of the binary. One of the aims of this work is to quantify the interaction between EMRI signals and these higher-order modes, and to assess their impact on parameter inference.

The \texttt{PhenomHM} waveform model for MBHBs is characterized by a set of intrinsic and extrinsic parameters as shown in Table~\ref{tab:mbhb_parameters}. The intrinsic parameters, $\{M_T, q, a_1, a_2\}$, describe the physical properties of the binary system, including its total mass, mass ratio, and the spins of the two black holes respectively. The extrinsic parameters, $\{D_L, \iota, \beta, \lambda, \psi\}$, determine the orientation and sky location of the source relative to the observer. In addition, $\phi_{\mathrm{ref}}$ specifies the reference orbital phase of the waveform, while $t_{\mathrm{ref}}$ defines a fine adjustment to the reference coalescence time. The full coalescence time is given by $t_{\mathrm{coal}} = t_0 + t_{\mathrm{ref}}$, where $t_0$ is a reference time within the LISA observing window. These parameters collectively determine the waveform amplitude, phase evolution, and detector response. To accelerate waveform evaluation, we compute the frequency-domain strain on a sparse grid of 1024 frequency points, reducing the computational cost compared to evaluating the waveform on the full frequency grid. We then interpolate the waveform from this sparse grid onto the full frequency grid used in the analysis. The interpolation is GPU-accelerated, enabling waveform generation in under 40 ms per evaluation. 
\begin{table}[ht]
\centering
\begin{tabular}{ll}
\hline
\textbf{Parameter} & \textbf{Description} \\
\hline

$M_T$ & Total binary mass \\
$q$ & Mass ratio \\
$a_1$ & Primary MBH spin \\
$a_2$ & Secondary MBH spin \\
$D_L$ & Luminosity distance \\
$\iota$ & Inclination angle \\
$\beta$ & Ecliptic latitude \\
$\lambda$ & Ecliptic longitude \\
$\psi$ & Polarization angle \\
$\phi_{\mathrm{ref}}$ & Reference phase \\
$t_{\mathrm{ref}}$ & Reference time  \\
\hline
\end{tabular}
\caption{Parameters describing the MBHB waveform in the PhenomHM model.}
\label{tab:mbhb_parameters}
\end{table} 
\section{Gravitational wave data analysis}
\subsection{Signal and Noise Description}
\label{analysis_and_noise}
As noted above, LISA data contain a superposition of many sources, resolvable and unresolvable, in addition to instrumental noise. We simulate EMRIs with FEW in the time domain and MBHBs with \texttt{PhenomHM} using \texttt{BBHx} in the frequency domain. The EMRI signals are sampled with a cadence of $\Delta t = 5$ seconds over an observation period of 1 year. The resulting waveform is then transformed into the frequency domain through a discrete Fourier transform. On the same frequency grid, we evaluate the MBHB waveform directly in the frequency domain and combine both signals together (with the simulated detector noise, when included). The final data stream in general can therefore be written as
\begin{equation}
    d(f) = h_{\mathrm{EMRIs}}(f) + h_{\mathrm{MBHBs}}(f) + n(f),
\end{equation}
where $n(f)$ denotes the detector noise realization and $h=h_{\mathrm{EMRIs}} + h_{\mathrm{MBHBs}}$ is the signal.

In this work, we model the detector noise as a weakly stationary Gaussian process with zero mean. Weak stationarity implies that the noise autocovariance function depends only on the time lag between two observations, rather than on the individual observation times. By the Wiener--Khinchin theorem \cite{Wiener1930,Khinchin1934}, the Fourier transform of the autocovariance function defines the power spectral density (PSD) of the noise. In GW data analysis, it is conventional to use the one-sided PSD, which is defined for positive frequencies (and a real noise process) and is twice the corresponding two-sided PSD.

Under the assumptions of stationarity and zero mean, the noise is uncorrelated across frequencies, with variance set by its PSD, so the noise covariance matrix is diagonal in the frequency domain~\cite{Finn:1992wt, https://doi.org/10.1111/j.2517-6161.1957.tb00242.x}. For $f,f' > 0$,
\begin{equation}
{
\Sigma(f, f') = \langle \tilde{n}(f)\,\tilde{n}(f')^{*} \rangle = \frac{1}{2}\delta(f - f')S_{n}(f')\,,}
\end{equation}
where $\langle\cdot\rangle$ is an ensemble average (given ergodicity) and $S_n$ is the noise PSD. If the noise is additionally Gaussian, the real and imaginary parts are then independently distributed as:
\begin{equation}
    \operatorname{Re}(\tilde{n}(f_i)) \sim \mathcal{N}\!\left(0,\; \frac{S_n(f_i)}{4\,\Delta f}\right)
\end{equation}
\begin{equation}
    \operatorname{Im}(\tilde{n}(f_i)) \sim \mathcal{N}\!\left(0,\; \frac{S_n(f_i)}{4\,\Delta f}\right)\,,
\end{equation}
except at Nyquist frequency and zero frequency where the frequency domain noise is purely real and the variance increases by a factor of 2~\cite{Burke:2025bun}. Under the assumption of stationary Gaussian noise, and excluding the zero-frequency (DC) and Nyquist frequency components, the log-likelihood can be written, up to an additive constant, as
\begin{equation}
\log \mathcal{L} =
-\frac{1}{2}\langle d-h \, | \, d-h \rangle
\label{eq:loglikelihood}
\end{equation}
where $h(t)$ denotes the analysis template used to model the underlying GW signal and we have assumed the noise PSD \(S_n(f)\) is perfectly known. Here, $\langle a|b\rangle $ defines the noise-weighted inner product between two signals in the frequency domain as:
\begin{equation}
\langle a \mid b \rangle
\simeq
4 \, \Delta f \,
\mathrm{Re}
\left[
\sum_{k}
\frac{\tilde{a}(f_k)\,\tilde{b}^{*}(f_k)}
{S_n(f_k)}
\right],
\label{eq:innerproduct}
\end{equation}
where $\tilde{a}(f)$ denotes the Fourier transform of $a(t)$, while $S_n(f)$ represents the PSD. For a signal $a(t)$ sampled over a total observation time $T$ with a cadence $\Delta t$, the discrete Fourier transform is given by
\begin{equation}
\tilde{a}(f_k) = \Delta t \sum_{n=0}^{N-1} a(t_n)\, e^{-2\pi i f_k t_n},
\label{eq:dft}
\end{equation}
where $f_k=k/T$ denotes the discrete frequency bins, $N=T/\Delta t$ is the number of samples, and  $\Delta f=1/T$ is the frequency resolution. 

To reduce spectral leakage associated with the finite observation time, windowing can be applied prior to the Fourier transform of the time-domain EMRI waveform. In this work, the analysis is performed under the assumption of stationary Gaussian noise, characterized entirely by its PSD $S_n(f)$. In realistic LISA observations, however, the noise is expected to deviate from stationarity. Data gaps, instrumental glitches, and unresolved foreground sources such as GBs and low SNR EMRIs can introduce non-stationary and non-Gaussian features into the data stream. Such effects may bias parameter estimation and invalidate the assumptions underlying the Whittle likelihood structure ~\cite{Burke:2025bun,Burke:2026rjq}.
 
For GW signals, the optimal SNR is defined as
\begin{equation}
\rho_{\mathrm{opt}} = \sqrt{\langle h \mid h \rangle}.
\end{equation}
Projecting the template waveform onto the data stream, we obtain the detection SNR using the analysis model $h$:
\begin{equation}
    \label{eq:detection_SNR}
\rho_{\mathrm{det}} =
\frac{\langle d \mid h \rangle}
{\sqrt{\langle h \mid h \rangle}} .
\end{equation}
By additionally normalizing with $\sqrt{\langle d \mid d \rangle}$, corresponding to the signal norm, we obtain the overlap as
\begin{equation}
\mathcal{O}(d,h) =
\frac{\langle d \mid h \rangle}
{\sqrt{\langle h \mid h \rangle}\sqrt{\langle d \mid d \rangle}} .
\label{eq:overlap}
\end{equation}

\subsection{TDI Channels and LISA Response}

Time Delay Interferometry (TDI) is a well-known technique for suppressing laser noise in the LISA data stream. It employs subtractions of various combinations of  delayed phase measurements from the spacecraft. This allows inference on the weak GW signals buried within instrumental noise \cite{Tinto:1999yr,Tinto:2002de,Tinto:2004wu}. For simplicity, we will assume a static configuration of the LISA instrument, with equal and constant arm lengths, thereby allowing the second generation Michelson-like TDI data streams $\{X,Y,Z\}$ to be expressed in the $\{A,E,T\}$ basis via a suitable transformation. For a static LISA configuration, the noise process among the $\{A,E,T\}$ channels is uncorrelated, yielding a diagonal covariance matrix facilitating the cheap evaluation of Whittle likelihood. The TDI response for EMRI is projected using \texttt{fastlisaresponse} \cite{Katz:2022yqe}, which uses GPU acceleration.
In this form, the total noise-weighted inner product from Eq.[\ref{eq:innerproduct}] becomes:
\begin{equation}
{
\langle a | b \rangle
=
\sum_{i = A,E,T}
\left\langle a_i \middle| b_i \right\rangle .
\label{eq:channelinnerproduct}
}
\end{equation}
\subsection{Bayesian Inference and MCMC Sampling}

We adopt a Bayesian framework in the analysis of GW signals. Using Bayes' theorem, the posterior distribution conditioned on the model hypothesis $\mathcal{H}$ can be written as
\begin{equation}
P(\theta | d, \mathcal{H}) =
\frac{P(d | \theta, \mathcal{H})\, P(\theta | \mathcal{H})}
{P(d | \mathcal{H})} ,
\label{eq:bayes}
\end{equation}
where $\theta$ represents the set of parameters to be estimated and $d$ is the observed data-stream. Here, $P(\theta | d, \mathcal{H})$ denotes the posterior distribution, $P(d | \theta, \mathcal{H})$ is the likelihood function, $P(\theta | \mathcal{H})$ is the prior distribution, and $P(d | \mathcal{H})$ is the evidence conditioned on the model $\mathcal{H}$. 

To sample the posterior, we use Markov chain Monte Carlo (MCMC). This work uses the \texttt{Eryn} sampler \cite{Karnesis:2023ras}, which builds on emcee~\cite{2013PASP..125..306F}, a MCMC framework implementing several advanced sampling techniques, including parallel tempering and multiple proposal (or \textit{moves}). Depending on the analysis, we use two or three temperatures, with the temperature ladder determined natively by \texttt{Eryn}, and adopt the default proposal kernel  \textit{stretch} move. The numbers of iterations and walkers are adjusted depending on the simulation in order to improve convergence while also maintaining reasonable computational costs. 

We initialize the walkers close to the true injected parameters, as the goal of this work is to characterize signal interactions and assess their effects on parameter recovery rather than to perform a blind search over the full parameter space. For each source parameter, we adopt a sufficiently broad prior centered on the injected value. The priors are chosen to be locally uninformative, such that the posterior is dominated by the likelihood and the maximum a posteriori (MAP) estimate is coincident with the maximum-likelihood estimate.
Given a set of posterior samples $\theta^{(i)} \sim p(\theta \mid d)$, we define the “best-fit parameters” as the MAP estimate, 
\begin{equation}
\theta_{\mathrm{MAP}} = \arg\max_{\theta} \, p(\theta  \mid d).
\end{equation}

In practice, we approximate the MAP estimate by the posterior sample with the largest log-posterior value. We quantify parameter-recovery errors using the normalized deviation of the MAP estimate from the injected value,
\begin{equation}
    \hat\Delta_i \equiv
    \frac{\theta_{i,\mathrm{MAP}}-\theta_{i,\mathrm{inj}}}{\sigma_i},
\end{equation}
where $\sigma_i$ denotes the standard deviation of the one-dimensional marginalized posterior for parameter $\theta_i$. For the zero-noise simulations considered in this work, we regard a parameter estimate as significantly deviated at the (1$\sigma$) level if the absolute normalized error exceeds unity. More generally, our primary focus is on comparing the errors across different inference methods rather than on the absolute magnitude of the deviations. Although these errors are reported for the one-dimensional marginalized posteriors, the sampling itself is typically performed in a $D$-dimensional parameter space, where $D$ corresponds to the total number of parameters across all sources being inferred.

To assess the convergence of the chains, several diagnostics are employed. In particular, we monitor the Gelman--Rubin statistic (PSRF), the effective sample size (ESS), requiring $\mathrm{ESS} \gtrsim 50$, and perform visual inspections of the chains and posterior distributions. 

Throughout the analysis, we compare two inference strategies: a sequential-fit approach, where individual signals are iteratively inferred and removed while treating other sources as noise, and a computationally intensive joint-fit approach that simultaneously infers multiple sources using a combined likelihood. 

\section{Characterizing multiple EMRIs}
 \label{just_EMRI}

This section characterizes deviations in source parameter inference due to EMRI self-confusion in the absence of other sources and instrumental noise. The goal is to minimize these deviations while keeping the computational cost minimal. We also examine the inter-source correlations between parameters of different EMRIs. To this end, we use a joint fit to sample over all EMRIs present in the data stream simultaneously. This approach should, in principle, yield small errors, since all signals are fit simultaneously and no additional noise assumptions are violated. However, the resulting parameter space is increasingly high-dimensional, making sampling with an MCMC algorithm computationally challenging.

We therefore compare these results against a sequential-fit approach, in which the loudest EMRI is estimated individually while the remaining EMRI signals are treated as confusion noise. The fitted signal is then subtracted from the data, and the procedure is repeated sequentially for the remaining sources. It is reasonable to expect that substantial overlap between sources could lead to errors in parameter estimation and induce strong correlations among parameters. However, as conjectured by \citet{Chua:2021aah}, self-confusion is expected to become important only for populations of approximately $10^{4}$ or more EMRIs. This motivates investigating whether, for the smaller number of individually resolvable EMRIs expected in a typical LISA data set, sequential inference can provide a computationally efficient alternative to a fully joint analysis without substantially degrading parameter recovery.

\subsection{Characterization of Two EMRIs}

We begin with the configuration comprising two EMRI signals in the absence of instrumental noise. We aim to quantify the deviation introduced by the sequential fit approach when the remaining EMRI signal acts as confusion noise. We use the EMRI parameters given in Table~\ref{tab:true_params_2_EMRI_noise} in Appendix A for the signal injection, denoted as \textbf{Case I}. The parameters are chosen such that the two EMRIs have SNRs of $80.51$ and $70.95$, respectively. Their corresponding eccentricities at the plunge are $0.15$ and $0.118$. We assume that the LISA observation window begins during the inspiral phase with initial dimensionless semi-latus recta $p_0=7.0$ and $p_0=8.0$. The corresponding last stable orbits (LSOs) occur at $p\lesssim3.4$ and $p\lesssim3.0$, respectively, and we terminate the waveforms close to these values. We do not model the subsequent plunge, merger, or ringdown.

\begin{figure}[tbh]
    \centering
    \includegraphics[scale = 0.37]{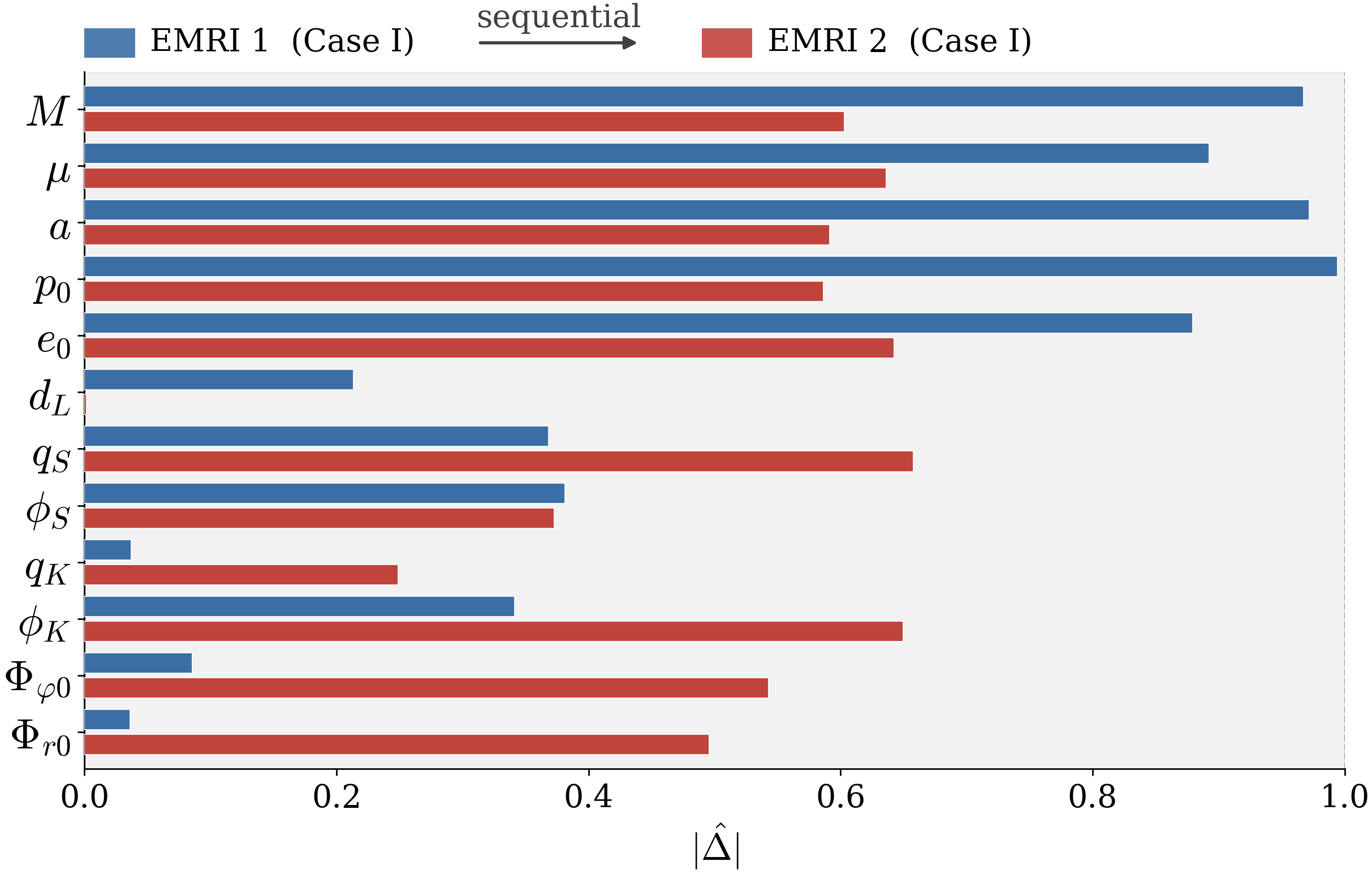}
    \caption{Normalized parameter deviations for data containing two EMRIs in the absence of instrumental noise. The louder source (EMRI 1) is inferred and subtracted first, followed by the characterization of the quieter source (EMRI 2).}
    \label{fig:emri_2_noise}
\end{figure}

For \textbf{Case I}, we perform sequential inference, characterizing the louder EMRI first while treating the second EMRI as confusion noise. After subtracting the best-fit waveform of the first EMRI from the data, we infer the parameters of the second EMRI from the residual. Figure~\ref{fig:emri_2_noise} shows the normalized parameter deviations for both sources, all of which remain within $1\sigma$ of their injected values. The blue bars represent the deviations for the louder EMRI, which is inferred first, while the red bars show those for the second EMRI.

Similarly, we analyze an additional \textbf{Case II}, in which the two EMRIs undergo coincident plunges. Such a configuration is very unlikely to occur in reality, but we include it for completeness and to test the robustness of the sequential fit. The parameters of the two EMRIs are given in Table~\ref{tab:true_params_2_EMRI_overlapping_AET}, Appendix A. The resulting SNRs for the two sources are $79.41$ and $69.40$, respectively. The corresponding eccentricities at the plunge are $0.152$ and $0.150$, consistent with the EMRI population models of \citet{Babak:2017tow}. We use both a sequential and a joint approach for this case. The joint fit also allows us to investigate the correlations between the parameters of the two EMRIs and assess whether this scenario introduces significant inter-source correlations.

\begin{figure*}[t]
    \centering
    \includegraphics[scale=0.55]{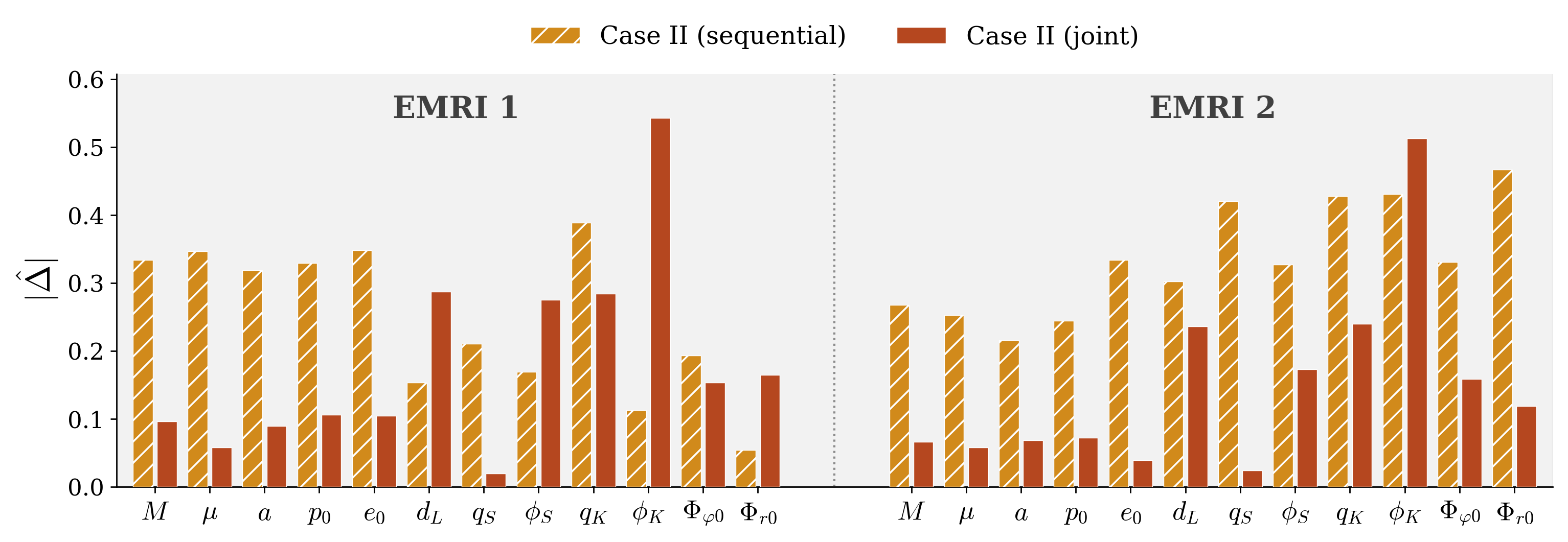}
    \caption{Two EMRIs with coincident plunges in the absence of instrumental noise. The normalized deviation is within $0.6\sigma$ for both sequential and joint fit.}
    \label{fig:bias_points_2EMRI}
\end{figure*}

Both approaches yield small normalized deviations for the one-dimensional marginals, with the absolute values within $0.6\sigma$. The comparison is shown in Fig.~\ref{fig:bias_points_2EMRI}, where absolute normalized errors for both approaches are presented. The errors from the joint approach are consistently smaller than those from the sequential fit, although the joint approach is substantially more computationally expensive.
\begin{figure}[h]
    \centering
    \includegraphics[scale = 0.42]{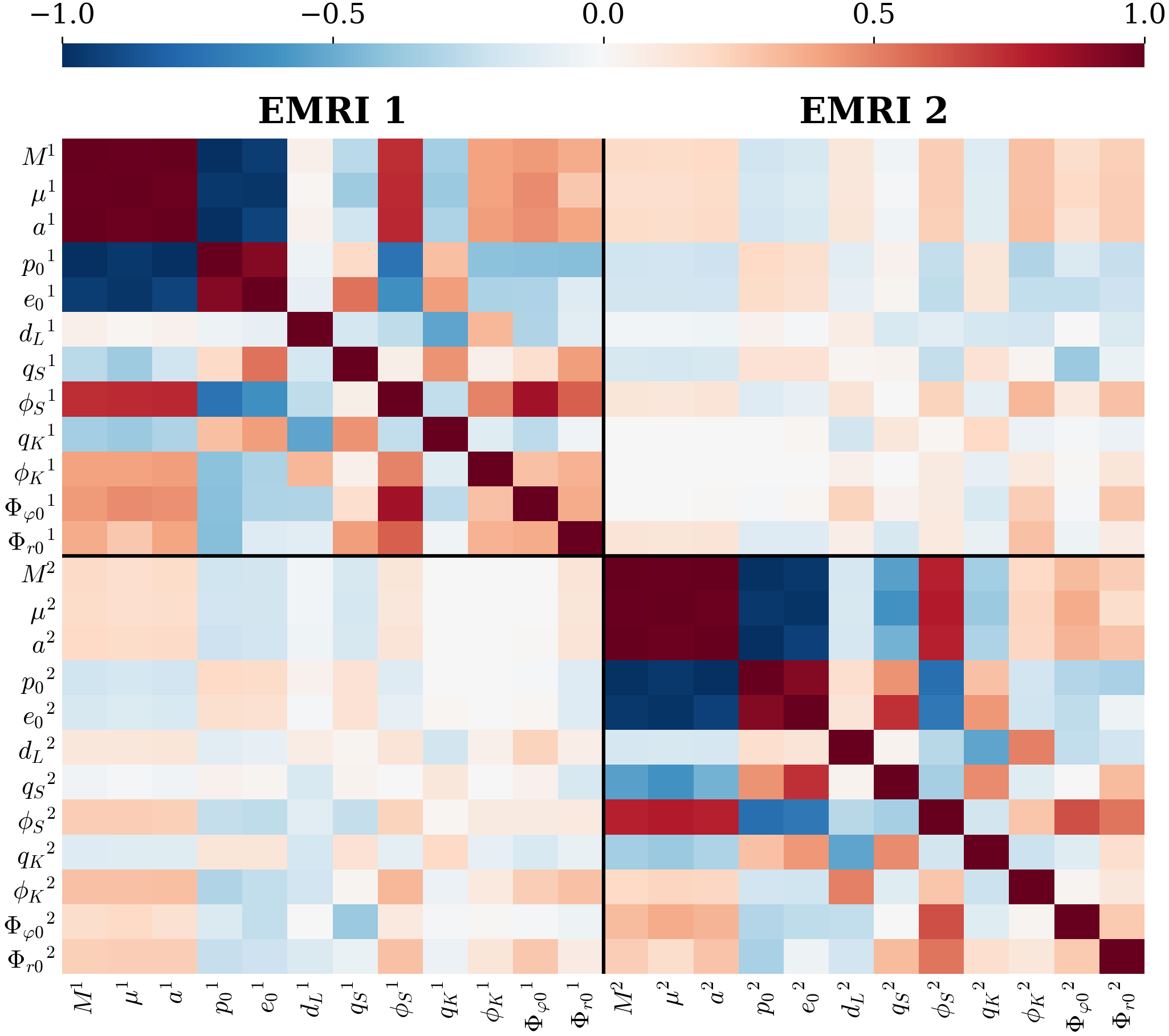}
    \caption{Correlation matrix - Simultaneous characterization of two EMRIs with coincident plunges. It depicts negligible inter-source parametric correlations even for a time-aligned plunging EMRI source.}
    \label{fig:corr_emri_2_overlapping}
\end{figure}
We also compute the correlation matrix, which describes the linear dependence between the parameters, defined as $C = \left(c_{i,j}\right)_{i,j\in\{1...N\}}$ :
\begin{equation}
    C_{ij} = \frac{\mathrm{Cov}(\theta_i, \theta_j)}{\sqrt{\mathrm{Var}(\theta_i)\mathrm{Var}(\theta_j)}}.
    \label{eq:corr_coeff2}
\end{equation}
We require the samples from the joint fit to evaluate the correlation matrix. The resulting correlation matrix is shown as a heat map in Fig. \ref{fig:corr_emri_2_overlapping}. It exhibits near-zero inter-source parameter correlations. This indicates that even when two EMRIs plunge at the same time, the correlations between their parameters remain negligible.

\subsection{Characterization of Three EMRIs}
 \label{3_EMRIs}
\begin{figure*}[t]
\centering
\includegraphics[width=\textwidth]{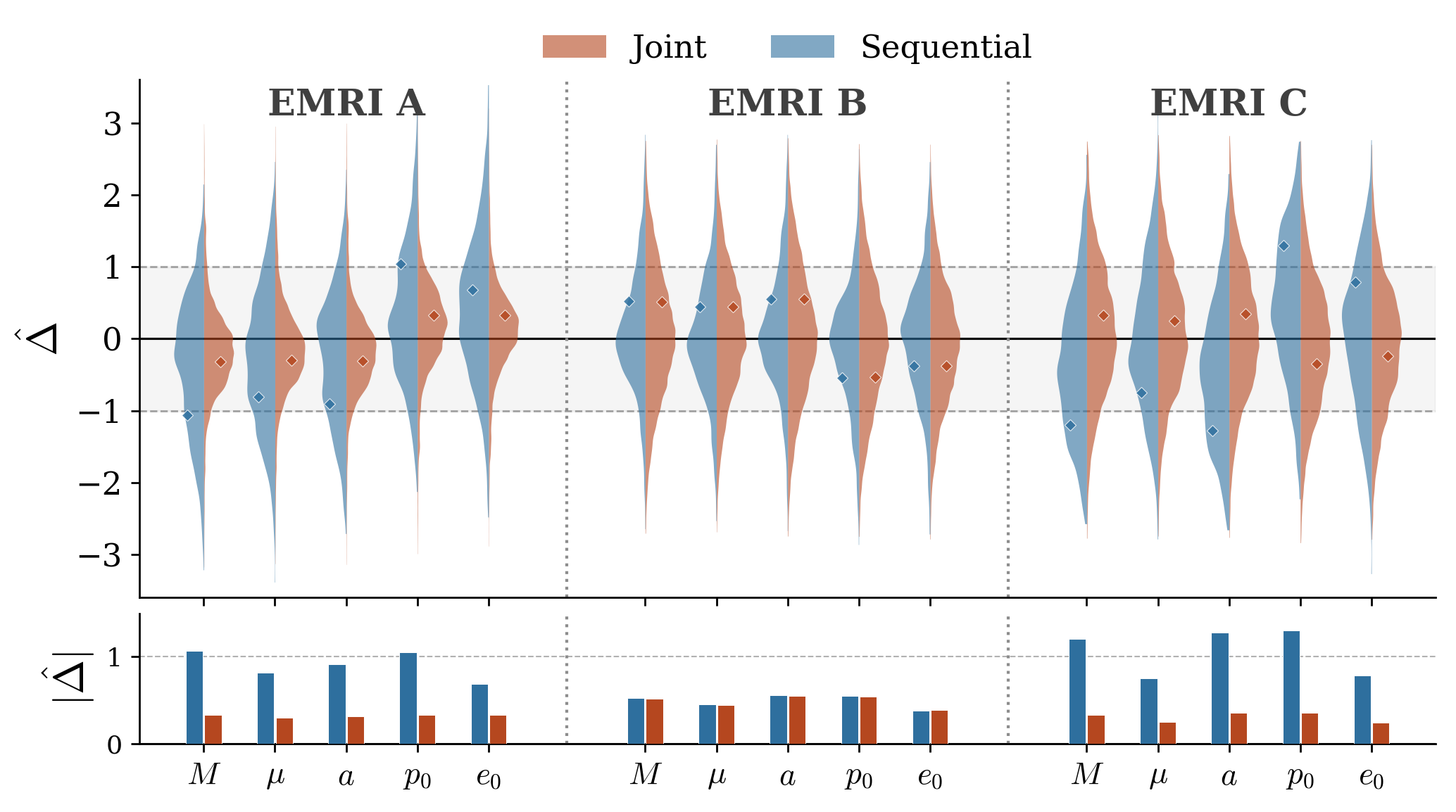}
\caption{Violin plots comparing the recovery of the intrinsic parameters of three EMRIs using the joint fit and sequential fit approaches. The joint fit demonstrates smaller normalized deviations but at a higher computational cost. In contrast, the sequential fit gives deviations near $1\sigma$ at substantially lower cost.}
\label{fig:combined_bias_global_vs_local2_3_EMRI}
\end{figure*}
\begin{figure}[h]
    \centering
    \includegraphics[scale = 0.37]{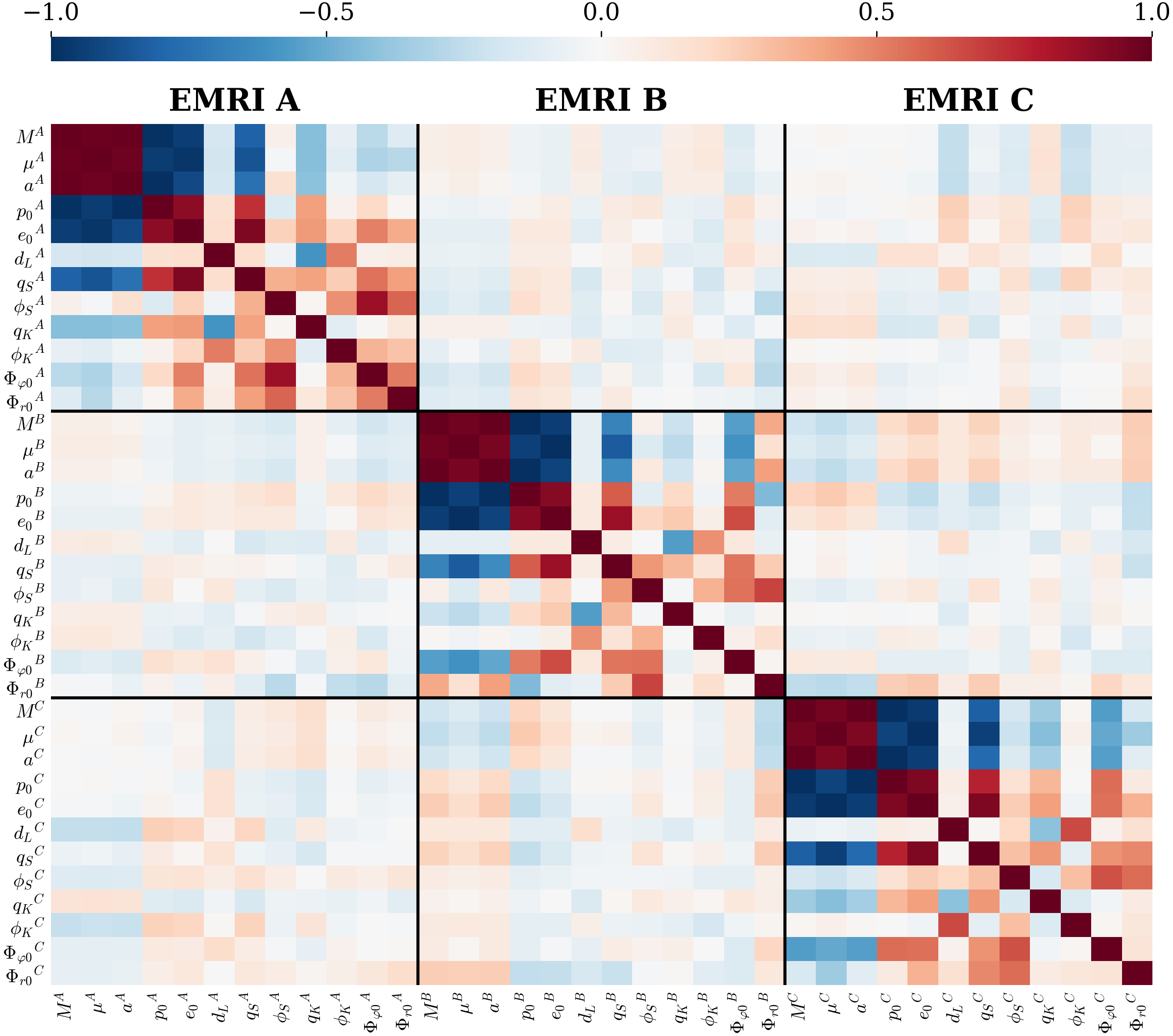}
    \caption{Correlation matrix - Simultaneous characterization of 3 EMRIs. It shows a near-zero correlation among the three EMRI signals we characterized using the joint fit approach.}
    \label{fig:3_EMRI_cross_source_correlation_250}
\end{figure}
\begin{table*}[htbp]
\centering
\caption{Injection parameters for the three EMRI signals}
    \label{tab:true_params_3_EMRI}
\renewcommand{\arraystretch}{1.15}
\begin{tabular*}{\textwidth}{@{\extracolsep{\fill}}lllllllllllll@{}}
\hline
\multicolumn{13}{c}{\textbf{EMRI Parameters for a 3-EMRI Datastream}} \\
\hline
Source & $M$ & $\mu$ & $a$ & $p_0$ & $e_0$ & $d_L$ (Gpc) & $q_S$ & $\phi_S$ & $q_K$ & $\phi_K$ & $\Phi_{\varphi0}$ & $\Phi_{r0}$ \\
\hline
EMRI A & 1007500 & 10.2500 & 0.8500 & 7.0000 & 0.1750 & 0.7500 & 0.7750 & 1.9500 & 1.5500 & 0.9500 & 1.9500 & 0.2750 \\
EMRI B & 1010000 & 10.5000 & 0.9000 & 7.5000 & 0.2500 & 0.8000 & 0.8000 & 2.0000 & 1.6000 & 1.0000 & 2.0000 & 0.3000 \\
EMRI C & 1012500 & 10.7500 & 0.9500 & 8.0000 & 0.3250 & 0.8500 & 0.8250 & 2.0500 & 1.6500 & 1.0500 & 2.0500 & 0.3250 \\
\hline
\end{tabular*}
\end{table*}
In this sampling configuration, we include three EMRI signals with near-zero pairwise overlaps. The data stream consists of one EMRI that plunges at $0.831$ years after the start of observation and two EMRIs that remain in the inspiral throughout the one-year observation period. The parameters of all three sources are listed in Table~\ref{tab:true_params_3_EMRI}. We refer to the plunging source as EMRI A and the remaining two sources as EMRI B and EMRI C. The corresponding SNRs are approximately $\rho_{\mathrm{EMRI_A}}=117$, $\rho_{\mathrm{EMRI_B}}=100$, and $\rho_{\mathrm{EMRI_C}}=71$. As in the previous configuration, the eccentricities at the plunge for all three sources lie in the range $0-0.2$. The pairwise overlaps between the EMRI waveforms, defined by Eq.~\ref{eq:overlap}, are as follows:
\[
\begin{aligned}
\mathcal{O}(h_{\mathrm{EMRI_A}}, h_{\mathrm{EMRI_B}}) &= -1.0 \times 10^{-2}, \\
\mathcal{O}(h_{\mathrm{EMRI_A}}, h_{\mathrm{EMRI_C}}) &= -8.5 \times 10^{-3}, \\
\mathcal{O}(h_{\mathrm{EMRI_B}}, h_{\mathrm{EMRI_C}}) &= 1.0 \times 10^{-4}.
\end{aligned}
\]
One might intuitively expect parameter estimation to be easier for signals with low overlaps, as the reduced similarity between the signals should limit inter-source degeneracies. This is particularly relevant given the low probability of observing three EMRIs with significant waveform overlap. Following the analysis presented in the previous subsection, we first perform a sequential fit and then a joint fit on the same signal configuration. Rather than displaying the full posterior distributions, we show the normalized one-dimensional marginalized posteriors of the intrinsic parameters for both approaches in Fig.~\ref{fig:combined_bias_global_vs_local2_3_EMRI}. For each parameter, the blue half-violin on the left corresponds to the sequential fit, while the brown half-violin on the right represents the joint fit. The diamond markers indicate the corresponding MAP estimates, and the lower panels show their absolute normalized deviations.

The joint fit exhibits small deviations, with all parameters remaining within 1$\sigma$ of their injected values. The sequential fit also recovers the injected parameters near 1$\sigma$, although the deviations are larger using this approach, as the other EMRI signals are treated as confusion noise. For a joint fit of three EMRIs, the combined parameter space has 36 dimensions. This requires at least 72 walkers and consequently leads to a substantial increase in computational cost. In our analysis, the joint fit requires approximately 10,000 iterations to reach convergence, using 80 walkers at three different temperatures. On a single A100 GPU, this requires nearly 200 hours of computation, compared with approximately 24 hours for the sequential fit. The convergence time also depends on the walkers' initialization, introducing an additional source of computational variability. As the number of EMRIs increases, the computational cost of a joint fit is therefore likely to become prohibitive. 

We initialize the walkers within small hyper-ellipsoids centered on the injected parameters. At the beginning of the sampling process, the walkers therefore occupy a small region of the parameter space relative to the eventual extent of the posterior. Consequently, empirical correlations estimated from the early samples can be strongly influenced by the initial configuration and may not reflect the underlying posterior correlations. As sampling progresses and the walkers explore the posterior, these correlations diminish, but at varying rates for different parameters, depending on the local structure of the posterior. Directions with stronger posterior curvature tend to constrain the walkers more rapidly, allowing the initial correlations to diminish within a few hundred steps. In contrast, weakly constrained directions with shallower posterior curvature can exhibit slower mixing, causing the ensemble to retain the imprint of its initial configuration for substantially longer. This is exactly the pattern we find for this case: $\Phi_{r0}$, $\phi_{K }$ and $qS$ take substantially longer to lose the imprint of their initial configuration and exhibit persistent inter-source correlations across the three EMRIs during the initial phases of the sampling. These parameters correspond to more weakly constrained directions of the posterior, which may explain their slower mixing.

The joint-fit samples allow us to construct the correlation matrix for the 3-EMRI configuration, as shown in Fig.~\ref{fig:3_EMRI_cross_source_correlation_250}. As expected, the parameters within each EMRI exhibit strong intra-source correlations, while the inter-source correlations are negligible, providing further motivation to adopt for EMRIs the sequential-fit approach or Gibbs-type sampling which is already used in the analysis of other LISA source types~\cite{Littenberg:2023xpl,Katz:2024oqg}.

\section{Characterizing Interactions Between an EMRI and an MBHB}
 \label{EMRI_MBH}
\begin{table*}[htbp]
\centering
\caption{Injected parameters for the MBHB and EMRI signals.}
\label{tab:true_params_1_MBH_1EMRI}
\renewcommand{\arraystretch}{1.15}
\begin{tabular*}{\textwidth}{@{\extracolsep{\fill}}llllllllllll@{}}

\hline
\multicolumn{12}{c}{\textbf{MBHB Parameters}} \\
\hline
$M_T$ & $q$ & $a_1$ & $a_2$ & $\iota$ & $D_L$ (Gpc) & $\phi_{\rm ref}$ & $\lambda$ & $\beta$ & $\psi$ & $t_{\rm ref}$ & \\
2025762 & 0.5367 & 0.6474 & 0.7388 & 0.5472 & 31.4025 & 0.7000 & 0.4425 & 0.8030 & 0.4736 & 8768750 & \\

\hline
\multicolumn{12}{c}{\textbf{EMRI Parameters}} \\
\hline
$M$ & $\mu$ & $a$ & $p_0$ & $e_0$ & $d_L$ & $q_S$ & $\phi_S$ & $q_K$ & $\phi_K$ & $\Phi_{\varphi 0}$ & $\Phi_{r0}$ \\
999500 & 9.7500 & 0.6950 & 9.0000 & 0.2500 & 0.7500 & 0.7750 & 1.9500 & 1.5500 & 0.9500 & 1.9500 & 0.2750 \\

\hline
\end{tabular*}
\end{table*}
In this section, we examine a noiseless data stream consisting solely of an EMRI and an MBHB signal. We aim to understand how these two signals interact during parameter inference and determine the most effective strategy for characterizing both sources.

When both signals are present in the data-stream and we search for the MBHB, the search statistics will selectively fit a burst-like signal whose SNR is concentrated around the merger (days to weeks). In contrast, the EMRI waveform behaves as a low-amplitude perturbation to the Gaussianity of the noise process. Conversely, when searching for the EMRI in the presence of the MBHB merger, the latter behaves as a very loud, short-duration glitch in the data. Even though the two waveforms remain nearly orthogonal, this degree of non-Gaussianity introduced by the MBHB can produce a significant deviation from the injected point in the EMRI source inference.

The Fisher matrix formalism  provides a useful framework for understanding the error estimates \cite{Flanagan:1997kp,Cutler:2007mi,Vallisneri:2007ev,Antonelli:2021vwg}. We can use the equations from \cite{Antonelli:2021vwg} to describe the error estimates as:
\begin{align}
    \Delta\theta^i_{\rm noise} &= (\Gamma^{-1})^{ij} \left( \partial_j h_a \,\middle|\, n \right) , 
    \label{eq:delta_theta_noise} \\
    \Delta\theta^i_{\rm conf} &= (\Gamma^{-1})^{ij} \left( \partial_j h_a \,\middle|\, \Delta H_{c} \right) ,
    \label{eq:delta_theta_confusion} \\
    \Delta\theta^i_{\rm sys} &= (\Gamma^{-1})^{ij} \left( \partial_j h_a \,\middle|\, \delta h \right) ,
    \label{eq:delta_theta_sys}
\end{align}
where $\Delta\theta^i_{\rm noise}$ is an error induced by instrumental noise, $\Delta\theta^i_{\rm conf}$ is the error induced by the unresolved confusion foreground, $\Delta\theta^i_{\rm sys}$ is systematic error arising from the mismatch between the true and analysis waveform. Recall that the Fisher matrix scales as $\Gamma_{ij} \sim \rho_a^2$, where $\rho_a$ is the SNR of the signal being characterized, while the inner product between the template derivative and a confusion source ($\Delta H_c$) scales as
$
\left( \partial_j h_a \,\middle|\,\Delta H_c \right) \sim \rho_a \, \rho_{\Delta H_c} \, \mathcal{O}_j ,
$ where we define
\begin{equation}
\mathcal{O}_j \equiv \frac{\left( \partial_j h_a \,\middle|\,\Delta H_c \right)}{\rho_a\, \rho_{\Delta H_c}}
\end{equation}
as the normalized inner product between the derivative of the template signal ($\partial_j h_a$) and the confusion source and $\rho_{\Delta_{H_c}}$ is the SNR of the confusion source. Substituting into Eq.~\eqref{eq:delta_theta_confusion}, the error induced due to confusion therefore scales as
\begin{equation}
    \Delta\theta^i_{\rm conf} \sim \mathcal{O}_j \, \frac{\rho_{\Delta H_c}}{\rho_a} ,
    \label{eq:bias_scaling}
\end{equation}
growing linearly with the SNR of the confusion source and suppressed by the SNR of the signal being inferred. Since EMRIs are expected to be much quieter than MBHBs, the confusion error for EMRI scales approximately as $\mathcal{O}_j\frac{\rho_{MBHB}}{\rho_{EMRI}}$. In contrast, when characterizing the MBHB in the presence of an EMRI, it scales approximately as $\mathcal{O}_j' \frac{\rho_{EMRI}}{\rho_{MBHB}}$. If we assume $\mathcal{O}_j\sim\mathcal{O}_j'$, and since $\rho_{\rm EMRI} \ll \rho_{\rm MBHB}$, the error in MBHB parameter inference will tend to be more strongly suppressed.

For the analyses that follow, we use the injected parameters as presented in Table~\ref{tab:true_params_1_MBH_1EMRI}. The resulting SNRs of an MBHB merger and an EMRI are 1169.76 and 55.39, while the overlap between these two sources is $9\times 10^{-4}$.  We analyze the same signal configuration using three different inference strategies:
\begin{enumerate}
\item \textbf{Case A:} A sequential fit in which the EMRI is characterized first. Its best-fit waveform is then subtracted from the data, after which the MBHB is inferred from the residual.
\item \textbf{Case B:} A sequential fit in which the MBHB is characterized first. Its best-fit waveform is then subtracted from the data, after which the EMRI is inferred from the residual.
\item \textbf{Case C:} A joint fit in which the MBHB and EMRI parameters are inferred simultaneously.
\end{enumerate}
\subsection{Sequential Inference with the EMRI First}

\begin{figure*}
    \centering
    \includegraphics[scale=0.62]{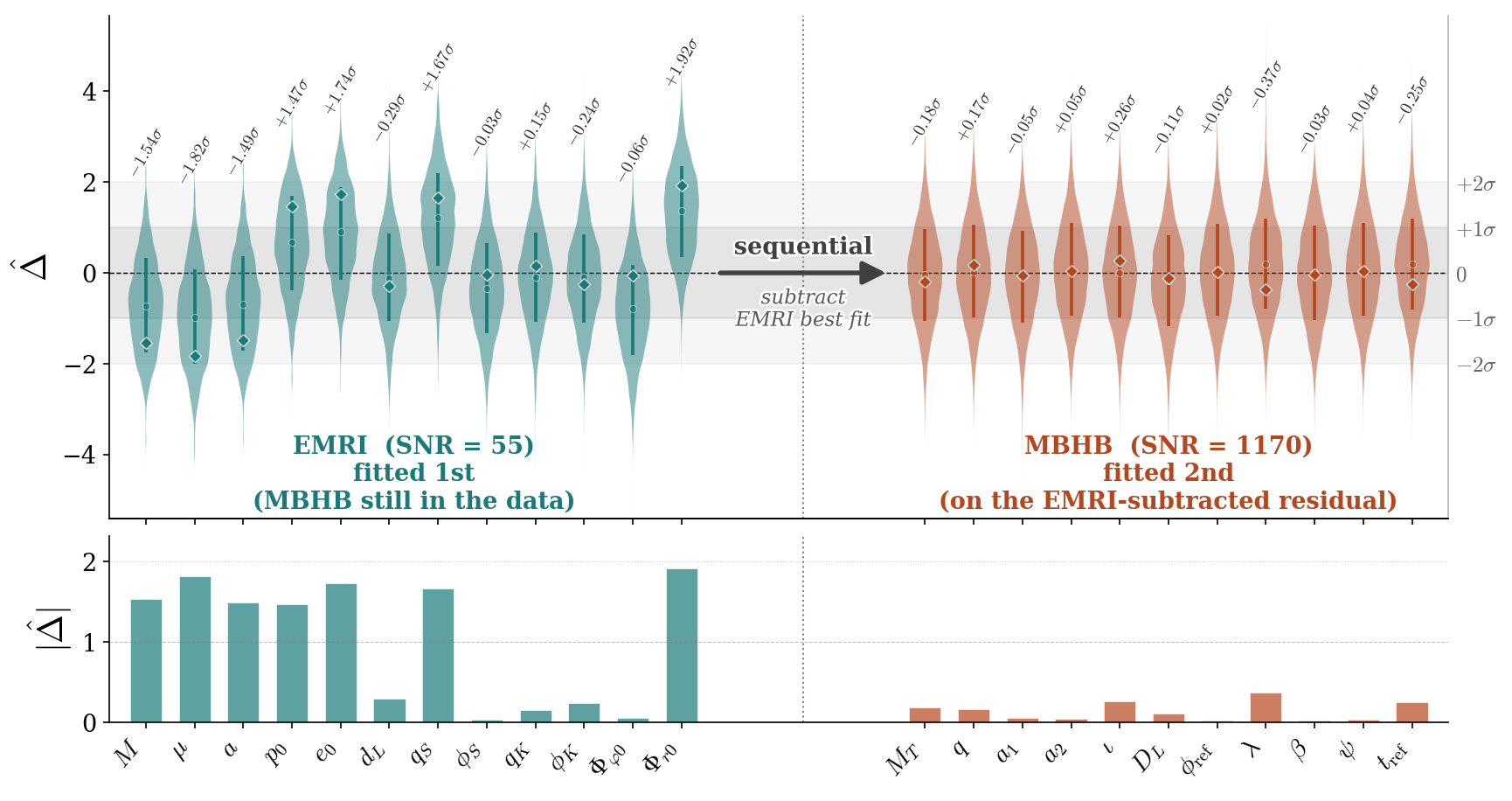}
    \caption{One-dimensional normalized marginals of the EMRI and MBHB source parameters, shown as violin plots. We first characterize the EMRI in the presence of the MBHB signal and subsequently infer the MBHB after subtracting the recovered EMRI signal. Some EMRI parameters exhibit normalized deviations exceeding the $1\sigma$ threshold, while the MBHB parameters show negligible deviations.}
    \label{fig:violin_bias_sequential_1_EMRI_1MBHB}
\end{figure*}

We begin by inferring the EMRI parameters while treating the MBHB signal as confusion noise for the configuration described above. The resulting violin plot is shown in Fig.~\ref{fig:violin_bias_sequential_1_EMRI_1MBHB}.  We recover significantly deviated estimates for several parameters, with normalized deviations exceeding the $1\sigma$ threshold. This suggests that the MBHB should be characterized before inferring the EMRI parameters. In practice, tens of MBHBs may be present in the LISA band, potentially increasing the impact on EMRI parameter recovery if they remain unmodeled.

Interestingly, despite the imperfect subtraction of the EMRI signal resulting from deviations in its recovered parameters, there is no significant downstream propagation of these errors into the subsequent MBHB inference. This asymmetry further motivates characterizing and subtracting the MBHB signal before proceeding with the EMRI inference.
\subsection{Sequential Inference with the MBHB First}

\begin{figure*}[t]
    \centering
    \includegraphics[scale=0.55]{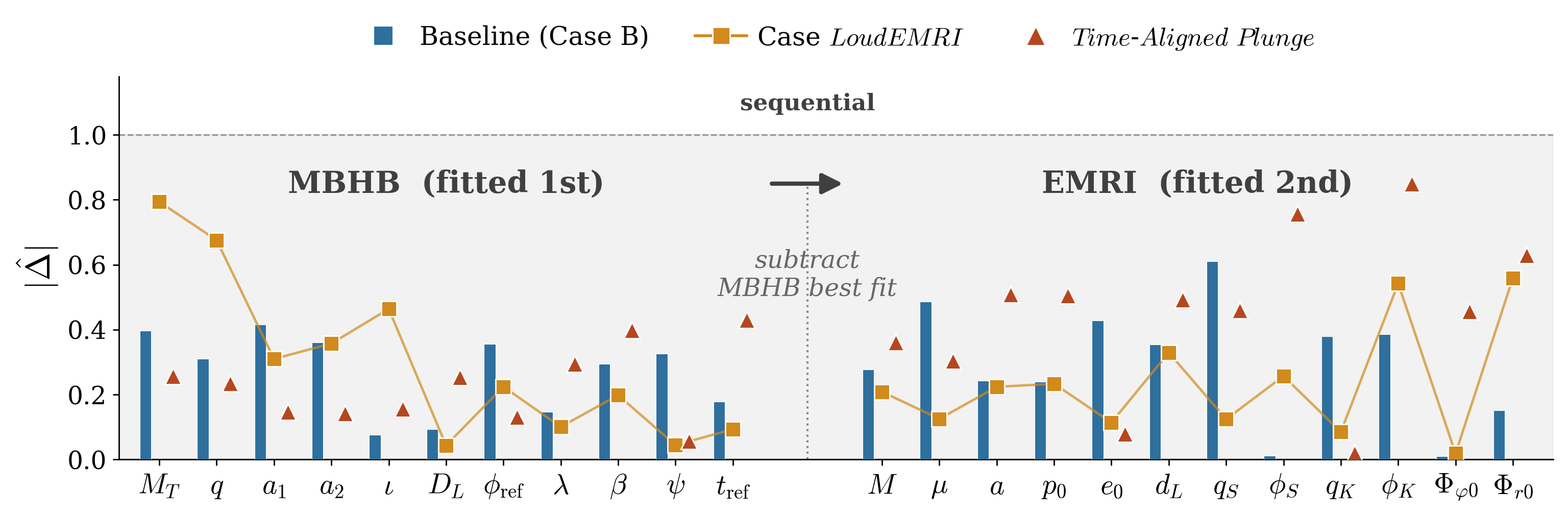}
    \caption{Deviations from the injected parameter values, normalized by the corresponding posterior standard deviations, for three cases analyzed using the sequential-fit approach. In each case, the MBHB is inferred first and its recovered waveform is subtracted from the data before inferring the EMRI parameters. The baseline case corresponds to the signal model in Table~\ref{tab:true_params_1_MBH_1EMRI}. In the \textit{LoudEMRI} case, the EMRI has a higher SNR than the MBHB, while the \textit{Time-Aligned Plunge} case corresponds to an EMRI plunge coinciding with the MBHB merger. For all three cases, the normalized deviations remain within 1$\sigma$.}
    \label{fig:bias_points_3configs}
\end{figure*}

The previous results suggest characterizing the MBHB first when the data contain both an EMRI and an MBHB. We now test the efficacy of this ordering across different configurations. We begin with the same system as in Table~\ref{tab:true_params_1_MBH_1EMRI}, but infer the MBHB before the EMRI. The results of this sequential fit are shown by the blue vertical bars in Fig.~\ref{fig:bias_points_3configs}, where all normalized parameter deviations remain within $1\sigma$ of their injected values.

To further assess the robustness of this strategy, we consider additional configurations. We examine a system containing a relatively low-SNR MBHB signal with $\text{SNR} = 516$ and a very loud EMRI signal with $\text{SNR} = 685$, which we denote as Case \textit{LoudEMRI}. The signal parameters are given in Table~\ref{tab:true_params_1_MBH_loud_EMRI}, Appendix A. Astrophysically, the SNR of an MBHB is expected to be much higher than that of an EMRI; however, we consider this configuration to investigate the regime in which the EMRI is louder than the MBHB. The two waveforms have an overlap of $4 \times 10^{-4}$.

For this configuration, we adopt the same sequential-fit strategy, inferring the MBHB first from a data stream containing both an EMRI and an MBHB. The normalized parameter deviations remain within $1\sigma$ for all parameters, as illustrated by the line plot in Fig.~\ref{fig:bias_points_3configs}. Interestingly, some MBHB parameters exhibit mild departures from Gaussianity, as seen in their posterior distributions shown in Fig.~\ref{fig:MBHB_characterization_loud_EMRI} in Appendix~B. Nevertheless, the MAP estimates remain within $1\sigma$ of their injected values.

Next, we analyze another EMRI--MBHB configuration in which the EMRI plunge time approximately coincides with the MBHB merger time. Although the EMRI signal builds up over an extended inspiral, a significant fraction of its SNR is still concentrated near the plunge. For the system considered here, the MBHB signal is exceptionally loud, with an SNR of 2397.82, while the EMRI has an SNR of 80.67. We denote this configuration as \textit{Time-Aligned Plunge}. The parameters are given in Table~\ref{tab:true_params_MBH_EMRI_overlapping}, Appendix A. 

\begin{figure}[h]
    \centering
    \includegraphics[scale=0.60]{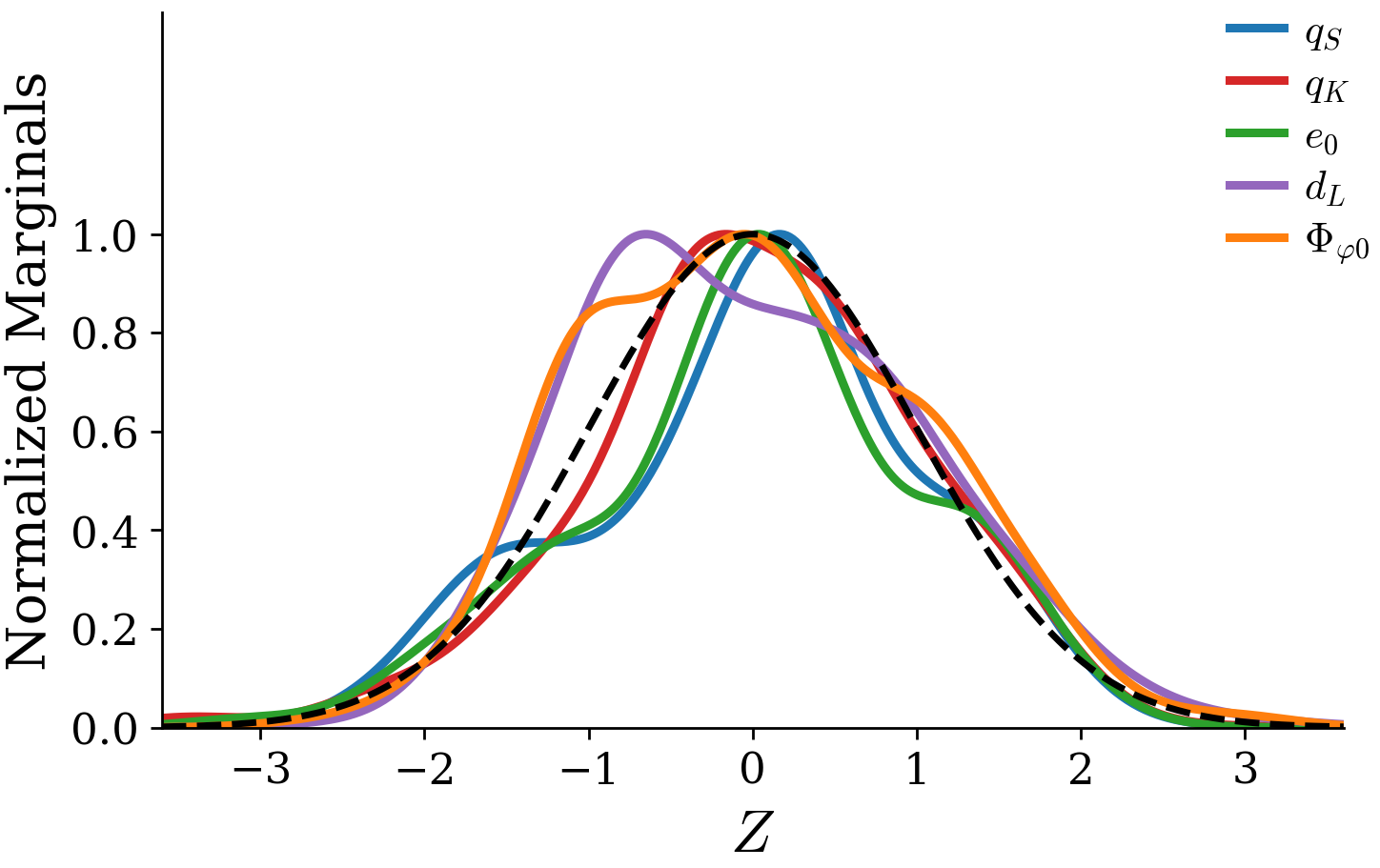}
        \caption{Normalized one-dimensional marginalized posteriors of EMRI parameters exhibiting significant departures from Gaussianity in the \textit{Time-Aligned Plunge} configuration, expressed in terms of their Z-scores. The dashed black line represents the standard normal distribution. The parameters shown are the initial eccentricity $e_0$, luminosity distance $d_L$, sky location polar angle $q_S$, initial BH spin polar angle $q_K$, and initial radial phase $\Phi_{r0}$.}
    \label{fig:nongaussianity_marginals_EMRI}
\end{figure}

The MBHB parameters are recovered without significant deviations. Although the normalized deviations of the EMRI parameters also remain small, their one-dimensional marginalized posteriors show noticeable departures from Gaussianity. These features are illustrated in Fig.~\ref{fig:nongaussianity_marginals_EMRI}, with the full posterior distributions shown in Fig.~\ref{fig:corner_plot_1emri_1mbhb_overlapping} in Appendix B. Importantly, these departures from Gaussianity do not translate into significant deviations from the injected parameter values, as illustrated by the scatter plot in Fig.~\ref{fig:bias_points_3configs}.
\begin{figure*}[t]
    \centering
    \includegraphics[scale=0.55]{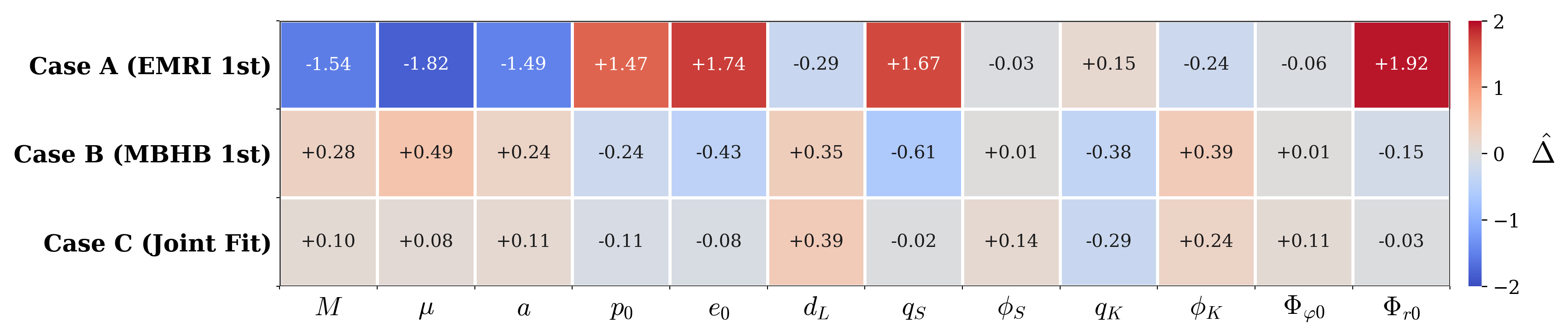}
    \caption{Normalized parameter deviations for the EMRI under three inference strategies: \textbf{Case A}, sequential inference with the EMRI characterized first. \textbf{Case B}, sequential inference with the MBHB characterized first; and \textbf{Case C}, joint inference of both sources. Overall, the deviations decrease from Case A to Case B to Case C.}
    \label{fig:bias_3run_EMRI_color}
\end{figure*}
\subsection{Joint Inference of an EMRI and an MBHB}

We next perform the joint inference introduced above as Case~\textbf{C}. Given the larger number of parameters, this approach is substantially more computationally expensive. Unlike the sequential approaches, however, the joint fit does not require one source to be treated as confusion noise while inferring the other, since both signals are modeled simultaneously. We therefore expect smaller deviations in the recovered parameters. 

The normalized deviations from the injected values remain within $1\sigma$ for all EMRI and MBHB parameters. As shown in Fig.~\ref{fig:bias_3run_EMRI_color}, the normalized deviations in the EMRI parameter estimates follow a clear ordering: they are smallest for the joint fit, intermediate for the sequential fit in which the MBHB is characterized and subtracted first, and largest when the EMRI is inferred in the presence of the unsubtracted MBHB signal.

\begin{figure}[h]
    \centering
    \includegraphics[scale = 0.35]{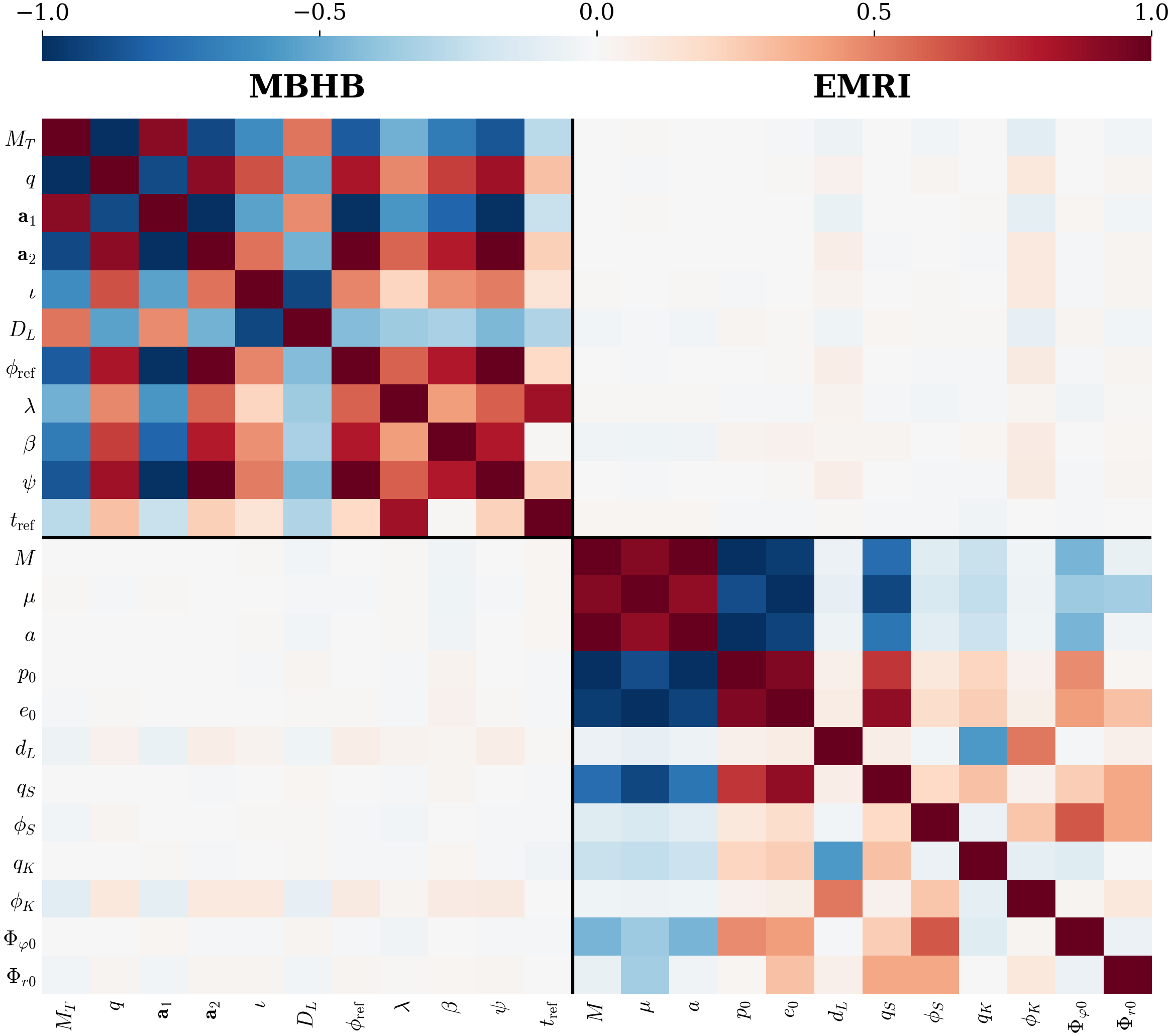}
    \caption{Correlation matrix obtained from the simultaneous characterization of an EMRI and an MBHB. The results show near-zero inter-source correlations between the EMRI and MBHB parameters.}
    \label{fig:corr_1_mbh_1_emri}
\end{figure}

To further understand the interaction between the two sources, we examine correlations among their parameters. Significant inter-source correlations would indicate that the characterization of the EMRI and MBHB is coupled, potentially making independent inference of the two sources more difficult. We use the posterior samples from the joint fit to construct the correlation matrix, with coefficients ranging between $-1$ and $1$. The resulting correlations are shown in Fig.~\ref{fig:corr_1_mbh_1_emri}. As expected, we find strong intra-source correlations among the parameters of both the MBHB and EMRI. 

In contrast, the inter-source correlations between the MBHB and EMRI parameters are nearly negligible. This weak coupling suggests that the two source classes can be sampled largely independently and provides further motivation for Gibbs-type sampling strategies for data streams containing multiple EMRIs and MBHBs.
\section{Characterization of Two MBHBs and an EMRI}
 \label{2MBH_1EMRI}

To extend our analysis to a more complex multi-source configuration, we introduce an additional MBHB signal into the data stream. Since most of the MBHB SNR is accumulated around the merger and the probability of two MBHBs merging simultaneously is expected to be low, we generally expect the overlap between MBHB signals to be small. In the following analysis, we investigate the parameter-recovery errors arising from the simultaneous presence of two MBHB signals and one EMRI signal in the absence of instrumental noise.

\paragraph{\textbf{Case O1}}: We choose the first MBHB to coalesce 10 days after the start of the observation with an SNR of $968.96$ and the second to coalesce after 241.5 days, with an SNR of $842.5$. We pair these sources with a non-plunging EMRI signal with an SNR of $98.56$. These values are representative of the SNRs expected for such systems. The corresponding parameters are given in Table~\ref{tab:MBH_2_EMRI_1_without_noise}, Appendix A. 

\begin{figure*}[t]
\centering
\includegraphics[width=\textwidth]{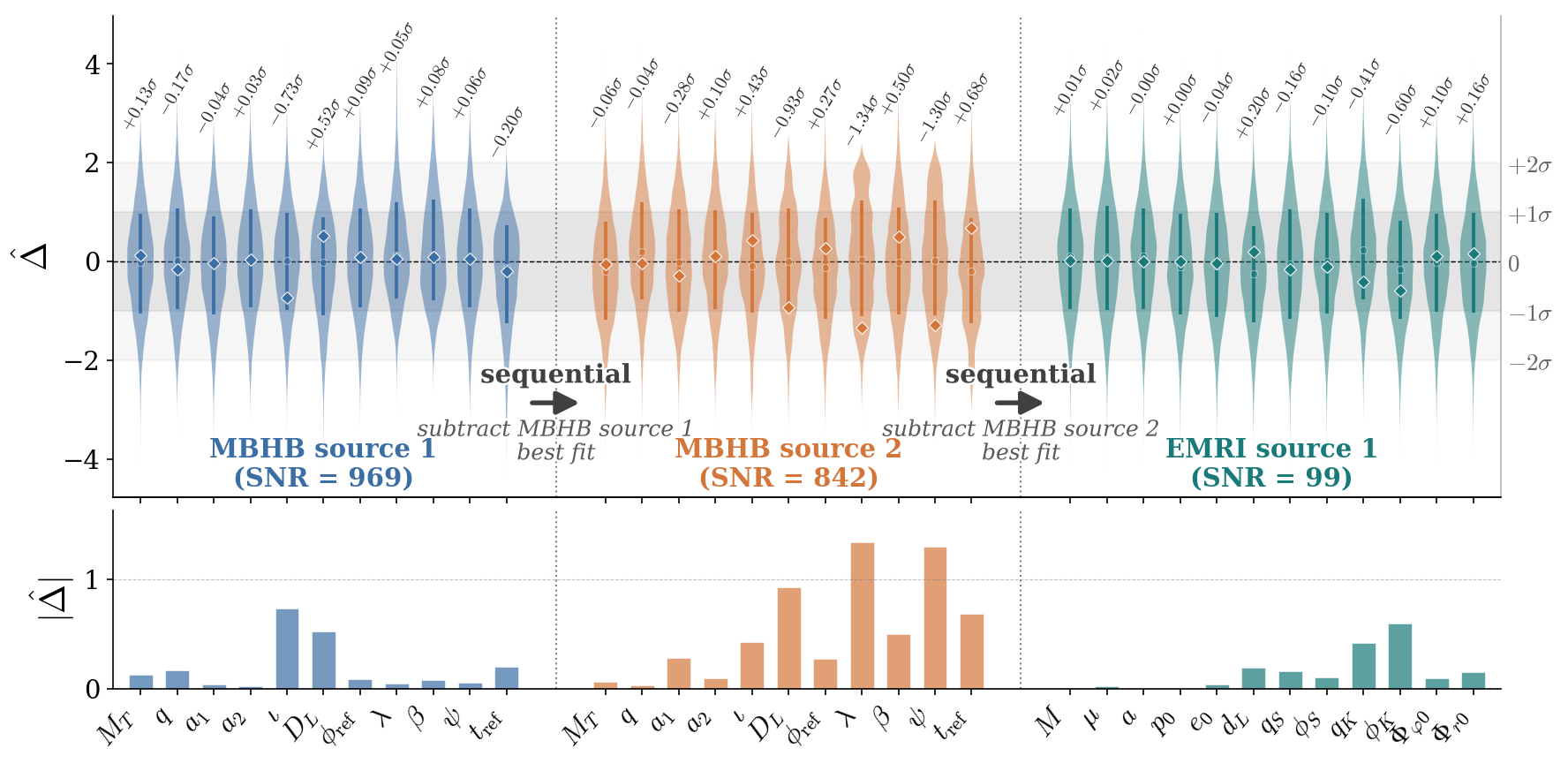}
\caption{Normalized posterior distributions and parameter deviations in units of $\sigma$ for the two MBHB and one EMRI system without the presence of noise. The parameter-recovery errors are around the $1 \sigma$ level.}
\label{fig:violin_bias_combined_2_MBH_1_EMRI}
\end{figure*}

We apply the sequential-fit approach to the three-source data stream, inferring the louder MBHB first in the presence of the second MBHB and an EMRI. After subtracting its best-fit waveform from the data, we infer the parameters of the second MBHB from the residual. We then subtract the recovered waveform of the second MBHB and characterize the EMRI from the resulting residual. The parameter-recovery errors are small, as shown in Fig.~\ref{fig:violin_bias_combined_2_MBH_1_EMRI}. The violin plot shows the normalized posterior distributions, with the diamond markers indicating the corresponding MAP estimates. Its lower panels show their absolute normalized deviations with a bar plot. These results support the use of sequential inference for this multi-source configuration. Next, we examine a more challenging case in which the two MBHB mergers are nearly coincident. Although such a configuration is expected to be uncommon, we include it to assess the robustness of sequential inference and quantify the resulting deviations in MBHB parameter recovery.
\paragraph{\textbf{Case O2}}: The signal consists of two MBHBs coalescing $2000$ seconds apart, with SNRs of $\rho_{\rm MBHB_1}\simeq1274.7$ and $\rho_{\rm MBHB_2}\simeq547.1$, together with an EMRI having an SNR of $\rho_{\rm EMRI}\simeq74.8$. The corresponding parameters are given in Table~\ref{tab:MBH_2_EMRI_1_overlapping_MBHs}, Appendix A. The two MBHB signals may be more difficult to resolve with sequential inference due to the close temporal proximity of their mergers. Nevertheless, the noise-weighted overlap between the two MBHB waveforms remains small, $\mathcal{O}(h_{\rm MBHB_1},h_{\rm MBHB_2}) =1.17\times10^{-3}$.  

First, we employ the sequential fit approach for all three sources as implemented in the previous subsection. We find that the MBHB parameters exhibit noticeable deviations, while the EMRI parameter recovery remains largely unaffected. To reduce the deviations in the recovered MBHB parameters, we jointly infer the two MBHB signals. We then subtract both recovered MBHB waveforms from the data before inferring the EMRI parameters from the residual.

\begin{figure*}[t]
\centering
\includegraphics[width=\textwidth]{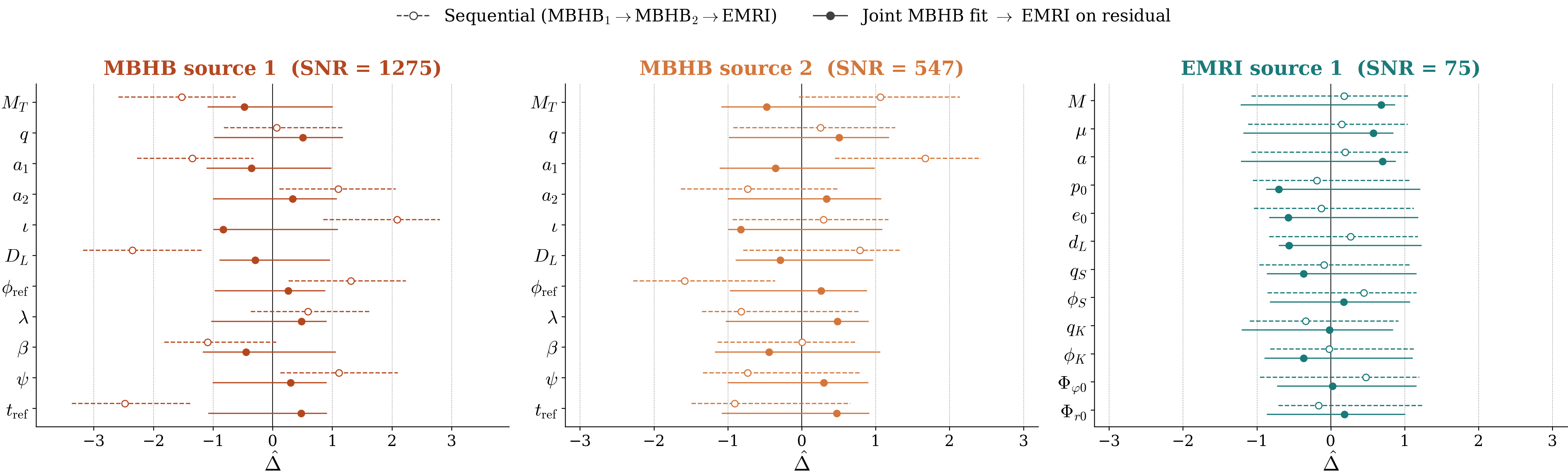}
\caption{Comparison of sequential and joint inference for two MBHBs with nearly coincident coalescence times. The dashed lines show the sequential approach, in which the two MBHBs are inferred separately, while the solid lines show the joint MBHB fit. In both approaches, the recovered MBHB signals are subtracted before performing EMRI parameter inference.}
\label{fig:bias_combined_2_MBH_overlapping_1_EMRI}
\end{figure*}

The corresponding error plots are shown in Fig.~\ref{fig:bias_combined_2_MBH_overlapping_1_EMRI}. It shows a reduction in parameter-recovery errors for the MBHBs using the joint MBHB fit compared with the sequential inference. Given the relatively low computational cost of MBHB likelihood evaluations, these results motivate jointly fitting MBHBs with temporally overlapping mergers before subtracting their recovered signals and subsequently performing EMRI inference.
\section{Systematic Effects of Missing Modes in MBHB Waveforms}
 \label{missing mode}
This section investigates the systematic biases induced by missing higher-order modes in the MBHB waveform model and the resulting impact of imperfect MBHB subtraction on EMRI parameter estimation. In LISA observations, MBHB signals generally contain contributions from multiple modes. However, due to computational limitations, only a restricted set of modes may be included in the waveform model used for parameter estimation. Although the omitted higher-order modes typically contribute less to the total SNR than the dominant modes, their absence may still introduce errors in the inferred MBHB parameters. Furthermore, imperfect subtraction of the MBHB signal can leave residuals that affect the subsequent recovery of the EMRI parameters.

We perform a proof-of-principle study of these effects by analyzing a data stream containing both an EMRI and an MBHB while deliberately omitting selected higher-order modes from the MBHB analysis waveform. Specifically, we consider two cases: one in which the $(4,3)$ mode is omitted and another in which both the $(4,3)$ and $(4,4)$ modes are omitted.

\subsection{Analysis with the $(4,3)$ Mode Omitted}
As described in Section \ref{MBHB_waveform_des}, the MBHB waveform is constructed using six modes, namely $(2,2)$, $(3,3)$, $(4,4)$, $(2,1)$, $(3,2)$, and $(4,3)$. The following analysis considers a signal containing one EMRI and one MBHB in the absence of instrumental noise, allowing us to isolate the effects of waveform-model mismatch without introducing additional stochastic uncertainty. The injected MBHB signal contains all six modes listed in Section \ref{MBHB_waveform_des}, whereas the analysis model ($h_m$) deliberately excludes the $(4,3)$ mode. This omission introduces a waveform mismatch between the injected signal and the model used during inference, which can lead to both a loss in detection SNR ($\rho_{\rm det}$) as defined in Eq.~\eqref{eq:detection_SNR} and deviations in the recovered parameters.

The impact of the missing mode can be understood geometrically in terms of the signal and analysis manifolds. The signal manifold comprises waveforms generated by the full waveform model without any omitted modes, whereas the analysis manifold is restricted to waveforms generated without the $(4,3)$ mode. The residual associated with the neglected $(4,3)$ mode can be decomposed into components parallel and orthogonal to the tangent space of the analysis manifold. The parallel component can be partially absorbed by variations in the model parameters, effectively compensating for the missing mode at the cost of shifting the inferred parameters. Conversely, the component orthogonal to the tangent space cannot be reproduced by varying the model parameters and therefore remains as an irreducible waveform mismatch. This orthogonal component primarily contributes to a loss in detection SNR and may also interfere with the characterization of other sources. In general, the resulting parameter deviations depend on the projection of the missing-mode residual onto the tangent space of the analysis manifold.

We quantify the significance of this effect using the SNR contribution of the $(4,3)$ mode, which varies across the MBHB parameter space. Consequently, the impact of omitting the $(4,3)$ mode is expected to vary between different MBHB configurations. This analysis therefore provides a controlled study of the consequences of an incomplete MBHB waveform model and how the resulting imperfect subtraction propagates into the inference of the accompanying EMRI signal. 

We consider two representative cases:

\textbf{Case S1:} The MBHB has a total SNR of $571.30$, a total mass of approximately $2\times10^6\,M_\odot$, and a mass ratio of $0.25$. The $(4,3)$ mode contributes an SNR of $19.90$. The accompanying EMRI has an SNR of $84.97$ and does not plunge during the observation.

\textbf{Case S2:} The MBHB has a total SNR of $1274.93$ and a total mass of $1.92\times10^6\,M_\odot$. In this configuration, the $(4,3)$ mode contributes an SNR of $8.70$, which is lower both absolutely and relatively than in Case S1. The accompanying EMRI has an SNR of $93.90$ and plunges approximately $250$ days into the observation. 

\begin{figure*}[t]
    \centering
    \includegraphics[width=\textwidth]{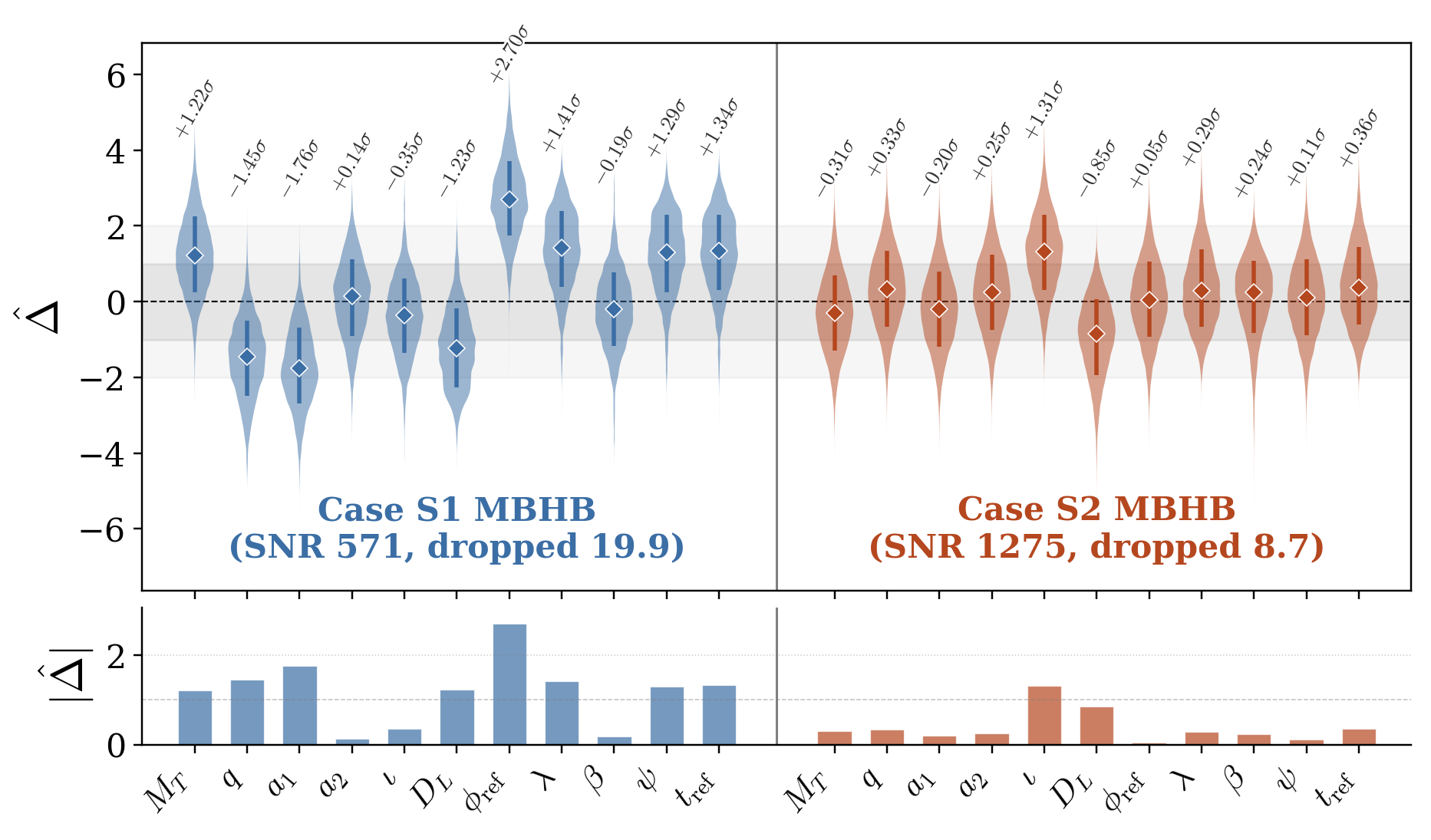}
    \caption{Normalized MBHB posteriors with the $(4,3)$ mode omitted. The \textit{S1} configuration exhibits larger parameter-recovery errors than \textit{S2}, consistent with the larger fraction of SNR lost by omitting this mode in \textit{S1}.}
    \label{fig:1_MBH_1EMRI_missing_mode_4_3_violin}
\end{figure*}

\begin{table}[htbp]
\centering
\renewcommand{\arraystretch}{1.2}
\begin{tabular}{lcc}
\hline
 & Case S1 & Case S2 \\
 & $M=2\times10^{6}\,M_\odot$ & $M=1.926\times10^{6}\,M_\odot$ \\
 & $q=0.25,\ \iota=75^\circ$ & $q=0.581,\ \iota=31^\circ$ \\
\cline{2-3}
Mode $(\ell,m)$ & SNR & SNR \\
\hline
$(2,2)$           & $511.8$ & $1268.6$ \\
$(3,3)$           & $213.0$ & $123.5$  \\
$(4,4)$           & $116.1$ & $46.7$   \\
$(3,2)$           & $47.9$  & $25.1$   \\
$(2,1)$           & $46.5$  & $18.5$   \\
$(4,3)$ (dropped) & $\mathbf{19.9}$ & $\mathbf{8.7}$ \\
\hline
Total (all modes) & $571.3$ & $1274.93$ \\
\hline
\end{tabular}
\caption{Per-mode SNR contributions for the two MBHB injections used in the missing-mode analysis, in which the $(4,3)$ mode is omitted from the MBHB inference template. Both configurations are noiseless, with $T_{\rm obs}=1\,\mathrm{yr}$. The mass ratio and inclination angle are denoted by $q$ and $\iota$, respectively.}
\label{tab:missing_43_snr}
\end{table}

The injection parameters for both configurations are listed in Table~\ref{tab:true_params_missing_43_compact} in Appendix~A, while the SNR contribution from each mode is given in Table~\ref{tab:missing_43_snr}. We analyze the system using a sequential-fit approach, in which the injected MBHB signal, which contains all six modes, is first characterized with an analysis waveform that omits the $(4,3)$ mode. After subtracting the corresponding best-fit waveform from the data, we infer the EMRI parameters from the residual.

The normalized one-dimensional marginalized posteriors of the MBHB parameters for both cases are shown in Fig.~\ref{fig:1_MBH_1EMRI_missing_mode_4_3_violin}. We do not show the EMRI posteriors, as we find no significant deviations in the inferred EMRI parameters in either configuration. The violin plots show that Case S1 exhibits larger deviations in the recovered MBHB parameters than Case S2. This is consistent with the larger contribution of the omitted $(4,3)$ mode in Case S1, although the magnitude of the parameter shifts also depends on the projection of the residual onto the model manifold.

We also quantify the impact of the missing $(4,3)$ mode on the recovered signal using the detection SNR, $\rho_{\rm det}$. Even after fitting the waveform without the $(4,3)$ mode, the recovered parameters largely compensate for the missing contribution, resulting in only a small loss in detection SNR. For Case S1, we obtain $\rho_{\rm det}=571.00$, corresponding to an SNR loss of $0.30$, while for Case S2, we obtain $\rho_{\rm det}=1274.90$, corresponding to an SNR loss of $0.03$. These results indicate that the omitted mode is partially absorbed through shifts in the inferred parameters, leaving only a small residual mismatch in the recovered waveform. Consequently, the residual has a negligible impact on the inference of the accompanying EMRI in these two configurations.
\subsection{Analysis with the $(4,4)$ and $(4,3)$ Modes Omitted}

\begin{figure}[h]
    \centering
    \includegraphics[scale = 0.45]{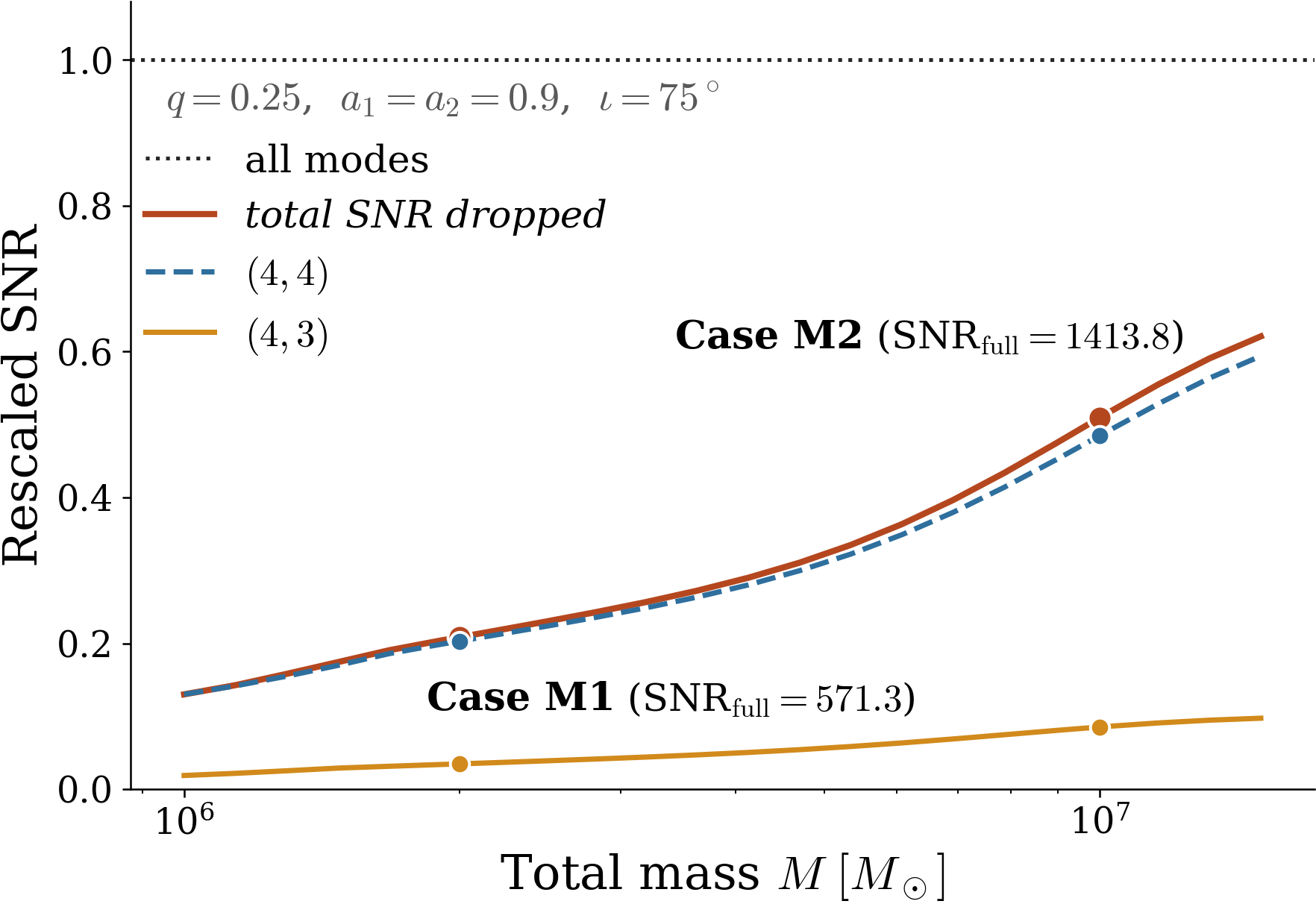}
    \caption{SNRs of the $(4,4)$ and $(4,3)$ modes, together with their combined contribution (\textit{total SNR dropped}), as a function of the total MBHB mass $M$. All other source parameters are held fixed at the values used in Cases \textbf{M1} and \textbf{M2}, except for the luminosity distance $D_L$, which is rescaled so that the full six-mode waveform has an SNR of unity. The circles mark the configurations corresponding to Cases \textbf{M1} and \textbf{M2}.}
    \label{fig:mass_scan_mode_snr}
\end{figure}

Similar to the analysis presented in the previous subsection, we consider a data stream containing one MBHB and one EMRI. In this study, however, the MBHB waveform used for inference omits two subdominant modes, $(4,3)$ and $(4,4)$, while both modes are retained in the injected signal. The relative SNR contributions of these modes vary significantly across the MBHB parameter space. In particular, as shown in Fig.~\ref{fig:mass_scan_mode_snr}, increasing the total mass of the MBHB leads to a more equitable distribution of SNR among the different modes, with the nominally subdominant modes contributing an increasingly significant fraction of the total SNR. We consider two representative MBHB configurations while keeping the EMRI configuration fixed in both cases. The EMRI does not plunge during the observation and has an SNR of $84.97$.
\begin{figure*}[t]
    \centering
    \includegraphics[width=\textwidth]{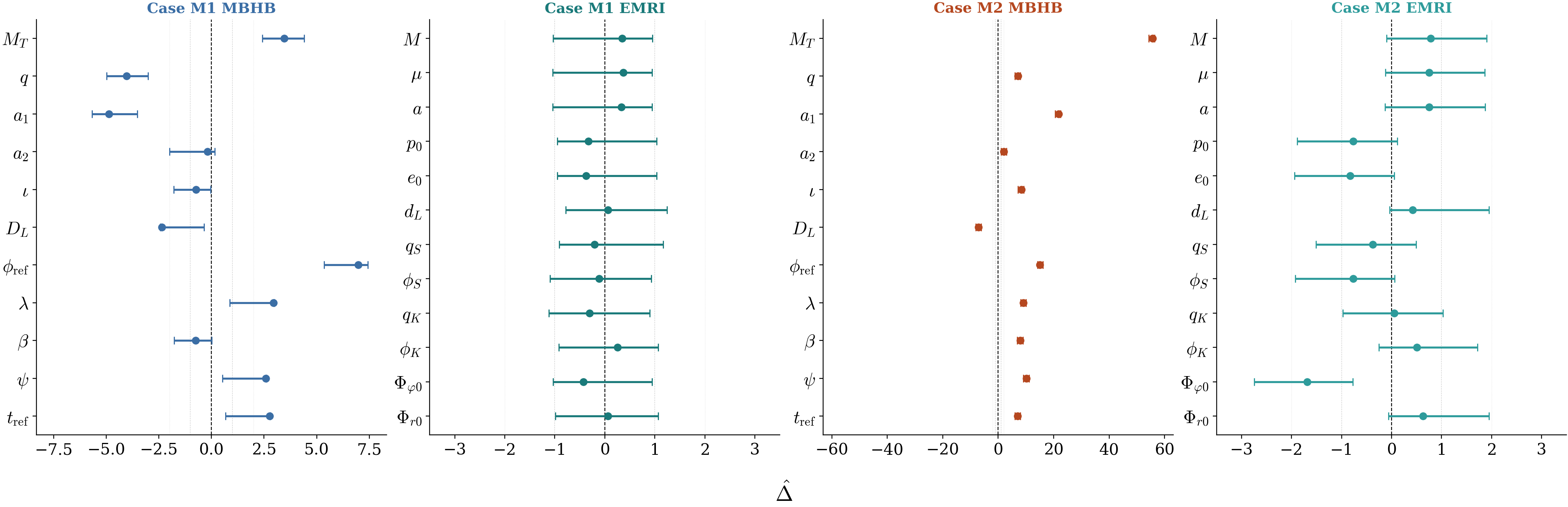}
    \caption{Error plot showing normalized parameter deviations when the $(4,3)$ and $(4,4)$ modes are omitted from the MBHB analysis waveform. For Case \textbf{M1}, the MBHB has a total SNR of $571.3$, with deviations in the recovered MBHB parameters reaching approximately $7\sigma$. For Case \textbf{M2}, the MBHB has a total SNR of $1413.8$ and exhibits larger deviations from the injected values, together with small deviations in the recovered EMRI parameters.}
    \label{fig:error_plot_all_44_43}
\end{figure*}

\textbf{Case M1:} We consider an MBHB with a total mass of $2\times10^6\,M_\odot$ and a mass ratio of $0.25$, yielding a total SNR of $571.3$. The $(4,3)$ and $(4,4)$ modes contribute SNRs of $19.9$ and $116.1$, respectively. This MBHB configuration is identical to Case S1 considered in the previous subsection, with the additional omission of the $(4,4)$ mode in the present analysis. The combined waveform contribution of the two omitted modes has an SNR of $119.2$.

We infer the source parameters sequentially, characterizing the MBHB first and then the EMRI after subtracting the recovered MBHB waveform. The omission of both modes produces substantial deviations in the recovered MBHB parameters, with some parameters deviating from their injected values by as much as $\sim7\sigma$. Despite these large errors in the MBHB inference, the EMRI parameters remain within $1\sigma$ of their injected values, as shown in Fig.~\ref{fig:error_plot_all_44_43}. This result is consistent with the relatively small impact of the missing modes on the recovered signal. After fitting, we obtain a detection SNR of $558.83$, corresponding to a modest SNR loss of $12.50$ relative to the injected SNR of $571.3$. Thus, although omitting both modes leads to appreciable deviations in the MBHB parameters, the resulting residual mismatch is sufficiently small that it does not significantly affect the inference of the EMRI parameters.

\textbf{Case M2:} To examine a configuration in which the omitted modes contribute a substantially larger fraction of the total signal power, we increase the MBHB total mass to $1\times10^7\,M_\odot$ while keeping the mass ratio fixed at $0.25$. At this higher total mass, the SNR contributions of the $(4,4)$ and $(4,3)$ modes increase substantially to $685.3$ and $120.7$, respectively. The combined contribution of the two omitted modes has an SNR of $720.2$. Consequently, omitting these modes is expected to produce a larger waveform mismatch and a correspondingly stronger impact on parameter inference. 

As in Case M1, we use sequential inference for source characterizations. As shown in Fig.~\ref{fig:error_plot_all_44_43}, we observe substantial deviations in the recovered MBHB parameters, with some parameter estimates displaced from their injected values by as much as $\sim55\sigma$.  In contrast to Case M1, we also find smaller but non-negligible deviations in the EMRI parameters. This behavior is consistent with the significantly larger amount of signal power excluded from the MBHB model when both the $(4,3)$ and $(4,4)$ modes are omitted. The resulting imperfect MBHB subtraction leaves a larger residual in the data, increasing the extent to which this residual can influence both the recovered MBHB parameters and the subsequent EMRI inference. This interpretation is further supported by the recovered detection SNR of $1221.56$, corresponding to an SNR loss of $192.20$ relative to the injected SNR of $1413.76$. Compared with Case M1, this substantially larger residual is consistent with the larger deviations observed in the EMRI parameters. We also find that the inferred total mass is systematically larger than the injected value in both cases.
\section{Analyzing Realistic LISA Data}
\label{lisa_data_analysis}
\begin{figure*}[t]
    \centering
    \includegraphics[width=\textwidth]{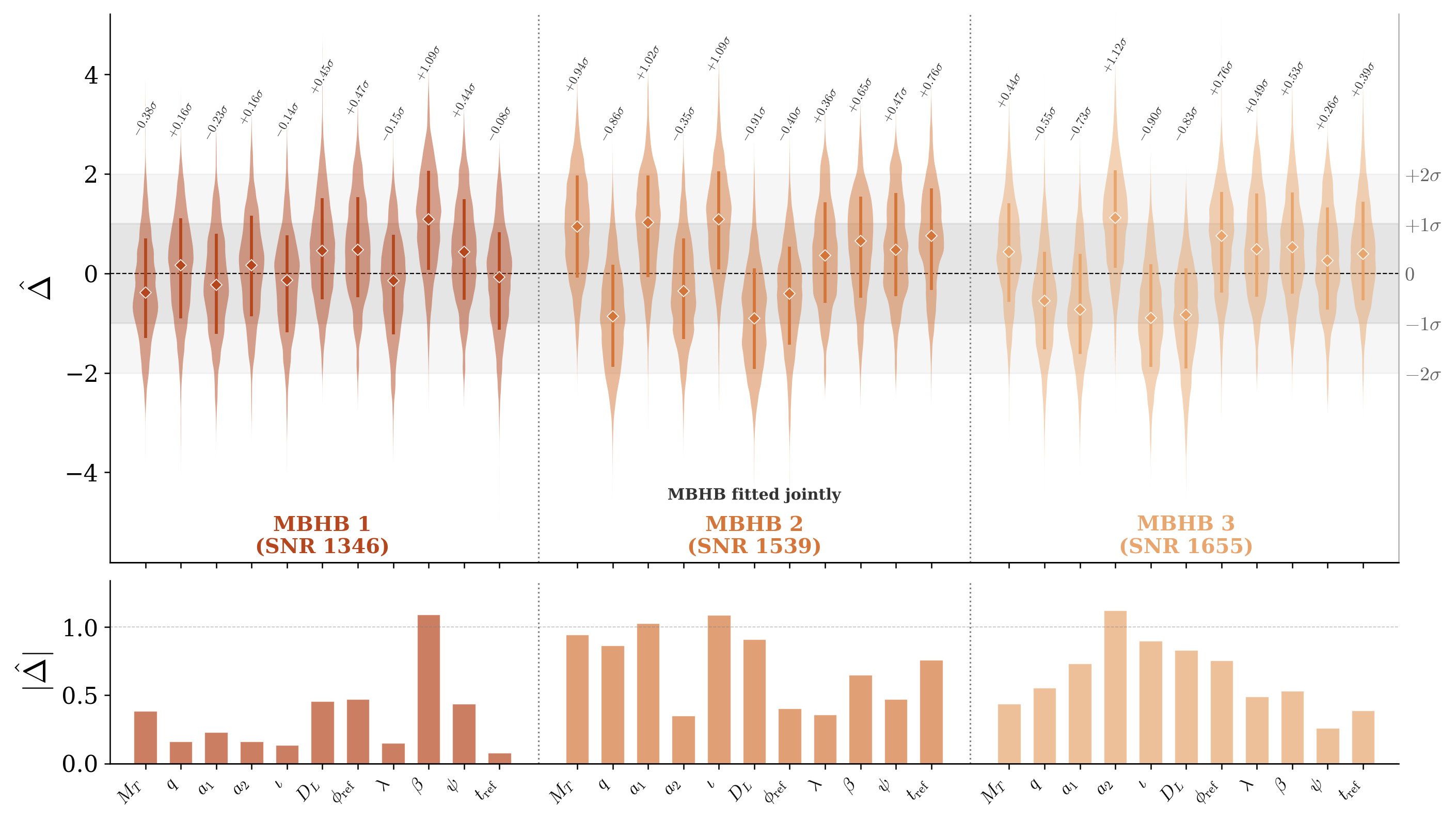}
    \caption{Normalized parameter deviations for the simultaneous characterization of three MBHBs in the presence of noise and three EMRI signals acting as confusion. The recovered deviations remain around the $1\sigma$ of their injected values.}
    \label{fig:violin_bias_3MBH_with_noise}
\end{figure*}
Having demonstrated the feasibility of sequential inference for EMRIs and joint inference for MBHBs, we now assess the performance of this strategy on a more realistic LISA data stream. We consider a data stream comprising three EMRIs and three MBHBs together with weakly stationary Gaussian noise generated as described in Section~\ref{analysis_and_noise}.

We generate three MBHB signals, three EMRI signals, and a noise realization sampled from the PSDs of the TDI channels. The adopted PSDs include the contribution from the unresolved white-dwarf confusion foreground. The source signals are generated using the injection parameters listed in Table~\ref{tab:true_params_3_MBH_3_EMRI}. The MBHB signals dominate the data stream and are therefore characterized first. We jointly infer their parameters and subtract the corresponding recovered waveforms from the data before proceeding with the characterization of the weaker EMRI signals. This ordering follows the results of the previous sections, where characterizing and subtracting the MBHBs first has been shown to reduce deviations in subsequent EMRI parameter recovery.

\subsection{Step 1: Joint fit approach on the MBHBs}

We select three MBHBs with total masses of around $2\times10^6\,M_{\odot}$ and mass ratios of around $0.65$, with corresponding SNRs of $1346$, $1539$, and $1655$. Since the MBHB signals may overlap in time, making them more difficult to resolve individually, we adopt a joint-fit approach. We simultaneously infer the parameters of all three MBHBs using an MCMC algorithm with $70$ walkers and two temperatures for $5800$ iterations, discarding the first $3350$ iterations as burn-in. The normalized one-dimensional marginalized posteriors of the MBHB parameters are shown as violin plots in Fig.~\ref{fig:violin_bias_3MBH_with_noise}. The diamond markers indicate the corresponding MAP estimates, while the lower panels show their absolute normalized deviations.

As shown in Fig.~\ref{fig:violin_bias_3MBH_with_noise}, the normalized deviations are small, with the injected parameters lying within the corresponding credible regions. The observed deviations are consistent with the statistical fluctuations expected from the noise realization. We then generate MBHB waveforms using the recovered parameter estimates and subtract them from the data stream. The resulting residual contains the three EMRI signals and noise, together with small residuals from imperfect MBHB subtraction. This substantially reduces contamination from the much stronger MBHB signals before performing EMRI inference.
\begin{figure*}[t]
    \centering
    \includegraphics[width=\textwidth]{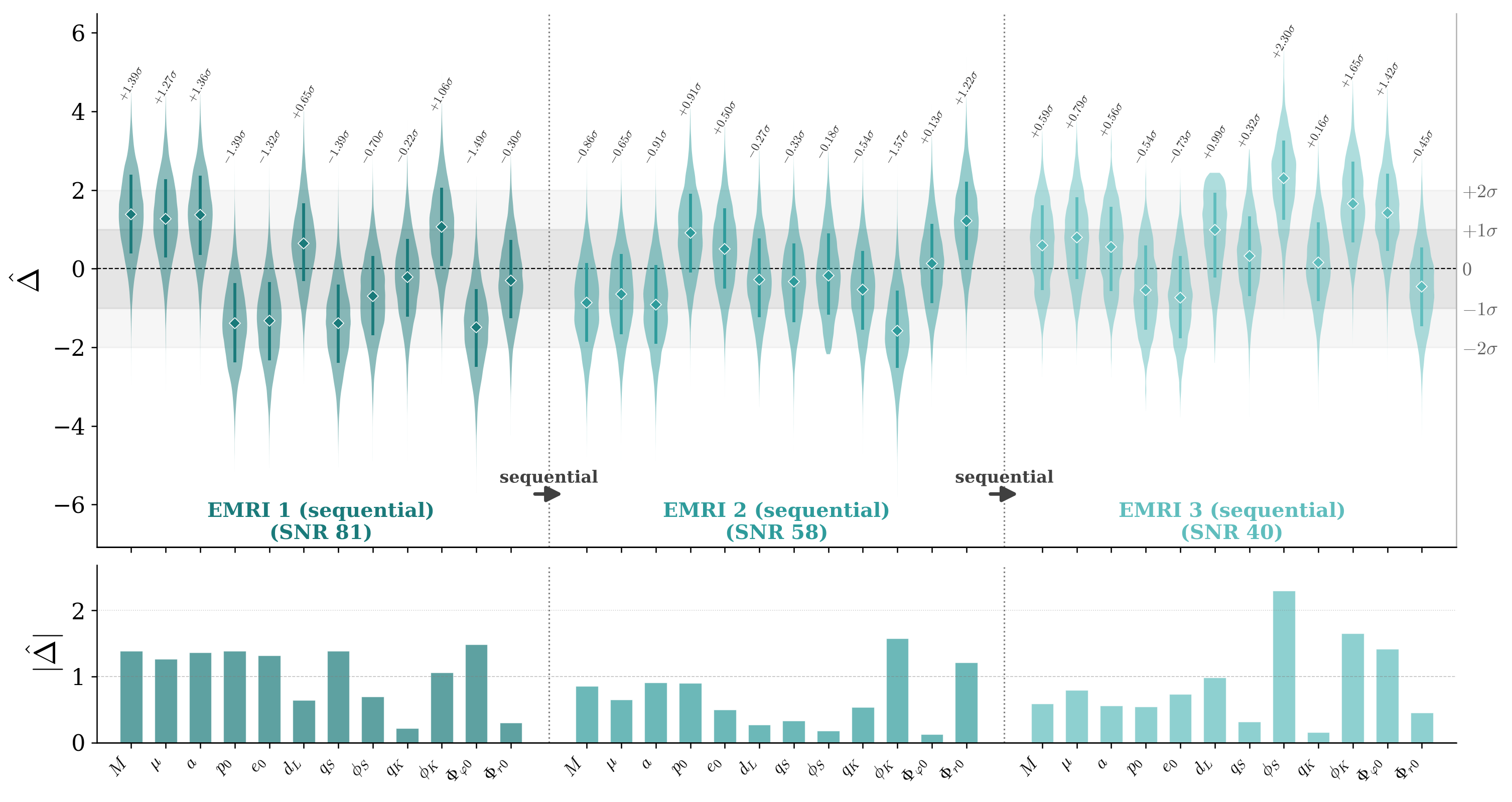}
    \caption{Normalized one-dimensional marginalized posteriors obtained from sequential inference of EMRI 1, EMRI 2, and EMRI 3 in the realistic LISA data stream after subtraction of the recovered MBHB signals. The resulting deviations remain within the corresponding credible regions, consistent with the expected statistical fluctuations due to noise.}
    \label{fig:violin_bias_3EMRI_with_noise}
\end{figure*}
\subsection{Step 2: Sequential fit on the EMRIs}
We consider three EMRIs, each with a primary mass of approximately $10^6\,M_{\odot}$ and a secondary mass of approximately $10\,M_{\odot}$. Their plunge eccentricities are $0.121$, $0.081$, and $0.097$, respectively, consistent with the population estimates of \citet{Babak:2017tow}. The corresponding SNRs are $81$, $58$, and $40$. Rather than performing simultaneous parameter estimation for the three EMRIs, we characterize them independently by running three sequential MCMC runs. In the first run, EMRI1 is inferred while treating EMRI2, EMRI3, and the noise as the background. In the second run, EMRI2 is characterized using EMRI3, the instrumental noise, and the residual from the EMRI1 characterization as the background. Finally, in the third run, EMRI3 is inferred using the instrumental noise and the residuals from the preceding inference steps as the background. Based on the results of Section~\ref{3_EMRIs}, we expect inter-EMRI contamination to be small for these configurations.

For each EMRI, we run an MCMC with 30 walkers for 5800 iterations, discarding the first 3350 iterations as burn-in. The resulting posterior distributions are shown as normalized violin plots in Fig.~\ref{fig:violin_bias_3EMRI_with_noise}. The diamond markers indicate the corresponding MAP estimates, and the lower panels show their absolute normalized deviations. The recovered parameters exhibit only small deviations from their injected values and remain within the corresponding credible regions, consistent with the expected statistical fluctuations due to noise. These results demonstrate that sequential characterization successfully recovers the parameters of all three EMRIs despite the simultaneous presence of multiple signals.

Overall, the results presented in this section demonstrate that the two-step strategy, consisting of joint inference of the MBHBs followed by sequential inference of the EMRIs, provides a viable and computationally efficient approach. The recovered posterior distributions contain the injected values of all source parameters within their corresponding credible intervals, with the observed deviations consistent with statistical fluctuations due to noise. Importantly, this is achieved without simultaneously inferring the parameters of all six sources, which would require a fully joint fit and incur a substantially higher computational cost. The proposed approach therefore reduces the computational burden while retaining reliable parameter recovery for the sources present in the data stream.

\section{Conclusion}

Our central result is a practical recipe for EMRI and MBHB inference in LISA data: fit the MBHBs first, jointly when their mergers are close in time, subtract them, then infer the EMRIs one at a time from the residual. In every configuration we studied, this recovers all source parameters within their credible regions, at a fraction of the cost of a fully joint fit. 

First, we consider data streams containing multiple EMRIs. For the configurations studied, the EMRIs can be characterized sequentially without substantial degradation in parameter recovery. The corresponding correlation matrices show negligible inter-source correlations between parameters belonging to different EMRIs supporting sequential inference as a computationally efficient alternative to a fully joint EMRI analysis.

For data streams containing both EMRIs and MBHBs, our results suggest that the MBHBs should generally be characterized and subtracted before proceeding with EMRI inference. For a year-long signal containing one MBHB and one EMRI, we demonstrate that the two sources can be inferred sequentially if the MBHB is characterized and subtracted first, followed by inference of the EMRI from the residual. For the configurations considered, the presence of the EMRI has a negligible impact on MBHB parameter recovery, whereas an unmodeled MBHB can produce significant deviations in the recovered EMRI parameters.

We find that the MBHB-first strategy remains effective in more challenging configurations, including cases in which the EMRI plunge approximately coincides with the MBHB merger and in which the EMRI has a higher total SNR than the MBHB. Using joint inference, we also find negligible inter-source correlations between the EMRI and MBHB parameters for the specific configurations considered. We expect this weak-correlation structure to remain qualitatively similar as the number of sources in these two classes increases, although verifying this expectation will require substantially greater computational resources. 

For signals containing multiple MBHBs and EMRIs, when two MBHB mergers occur close together in time, sequential MBHB inference can introduce noticeable deviations in the recovered parameters. In such cases, we find that joint inference of the MBHBs is preferable. Combining these results motivates a hybrid strategy in which temporally overlapping MBHBs are jointly characterized and subtracted, followed by sequential inference of the EMRIs. 

We additionally investigate the impact of omitting higher-order modes from the MBHB analysis waveform and the subsequent effect of imperfect MBHB subtraction on EMRI inference. For the configurations considered, missing higher-order modes can produce substantial deviations in the recovered MBHB parameters without necessarily producing comparable deviations in the EMRI parameters. In particular, for the $(4,3)$ and $(4,4)$ modes, part of the missing-mode residual can be absorbed through shifts in the recovered MBHB parameters when it projects onto the tangent space of the analysis-model manifold. Consequently, the loss in recovered detection SNR can be considerably smaller than the SNR associated with the omitted waveform contribution evaluated at the injected parameters. The component that cannot be absorbed remains as a residual after subtraction and is therefore more relevant for subsequent EMRI inference. Our results thus indicate that large deviations in the recovered MBHB parameters do not necessarily imply large deviations in the EMRI parameters.

Finally, we apply the combined inference strategy to a more realistic LISA data stream containing three MBHBs, three EMRIs, and a noise realization whose PSD includes the unresolved white-dwarf confusion foreground. We jointly characterize and subtract the MBHBs before sequentially inferring the EMRIs. The recovered parameters remain consistent with their injected values within the corresponding credible regions, demonstrating the feasibility of this approach for a more complex multi-source data stream without requiring a fully joint fit of all six sources.

It is important to emphasize that the results presented here constitute a proof-of-principle study rather than a population-level analysis. We consider only a limited set of representative EMRI and MBHB configurations, as the high dimensionality of the combined parameter space makes an exhaustive exploration of source parameters and signal combinations computationally prohibitive. Future studies should stress-test our proposed strategy on broader astrophysical populations and more realistic LISA data. In particular, incorporating populations of both resolvable and unresolved white-dwarf binaries, together with realistic EMRI and MBHB populations, will provide a more stringent test of the proposed inference strategy. Such analyses should also investigate departures from the weakly stationary Gaussian-noise assumptions adopted here, including non-stationary foregrounds, data gaps, and instrumental artifacts.

\section*{Acknowledgements}

R.A.~acknowledges the support of the President’s Graduate Fellowship of the National University of Singapore. O.B.~acknowledges financial support from the Grant UKRI972 awarded via the UK Space Agency. This project has received
financial support from the CNRS through the MITI interdisciplinary programs. This work makes use of the Black Hole Perturbation Toolkit~\cite{BHPToolkit}.

\section*{Data Availability}
This work uses the publicly available packages \texttt{FastEMRIWaveforms} (FEW)~\cite{Chua:2020stf,Katz:2021yft,Chapman-Bird:2025xtd}, \texttt{BBHx}~\cite{Katz:2020hku}, \texttt{fastlisaresponse}~\cite{Katz:2022yqe} and \texttt{Eryn}~\cite{Karnesis:2023ras}. The analysis code and posterior samples
supporting this study will be made publicly available on Zenodo upon publication.
\clearpage
\appendix
\onecolumngrid 
\section{Signal Parameters}
\label{appendix_A}

\begin{table*}[htbp]
\caption{True parameters used for the MBHB and EMRI injections in the
loud-EMRI configuration (MBHB SNR $=515.7$, EMRI SNR $=684.8$; the EMRI is
louder than the MBHB).}
\label{tab:true_params_1_MBH_loud_EMRI}
\renewcommand{\arraystretch}{1.15}
\begin{tabular*}{\textwidth}{@{\extracolsep{\fill}}llllllllllll@{}}
\hline
\multicolumn{12}{c}{\textbf{MBHB Parameters}} \\
\hline
$M_T$ & $q$ & $a_1$ & $a_2$ & $\iota$ & $D_L$ (Gpc) & $\phi_{\rm ref}$ & $\lambda$ & $\beta$ & $\psi$ & $t_{\rm ref}$ \\

1945762 & 0.5841 & 0.8474 & 0.9388 & 1.5472 & 45.0000 & 1.7000 & 0.5425 & 1.0030 & 0.5736 & 10867620 \\
\hline
\multicolumn{12}{c}{\textbf{EMRI Parameters}} \\
\hline
$M$ & $\mu$ & $a$ & $p_0$ & $e_0$ & $d_L$ & $q_S$ & $\phi_S$ & $q_K$ & $\phi_K$ & $\Phi_{\varphi 0}$ & $\Phi_{r0}$ \\
1012500 & 10.7500 & 0.9500 & 7.7000 & 0.3250 & 0.1000 & 0.8250 & 2.0500 & 1.6500 & 1.0500 & 2.0500 & 0.3250 \\
\hline
\end{tabular*}
\end{table*}

\begin{table*}[htbp]
\centering
\caption{True parameters used for the coinciding MBHB merger and EMRI plunge (MBHB SNR $=2397.82$, EMRI SNR $=80.67$);
noiseless}
\label{tab:true_params_MBH_EMRI_overlapping}
\renewcommand{\arraystretch}{1.15}
\begin{tabular*}{\textwidth}{@{\extracolsep{\fill}}llllllllllll@{}}
\hline
\multicolumn{12}{c}{\textbf{MBHB Parameters}} \\
\hline
$M_T$ & $q$ & $a_1$ & $a_2$ & $\iota$ & $D_L$ (Gpc) & $\phi_{\rm ref}$ & $\lambda$ & $\beta$ & $\psi$ & $t_{\rm ref}$ \\
2025762 & 0.5367 & 0.6474 & 0.7388 & 0.5472 & 31.4025 & 0.7000 & 0.4425 & 0.8030 & 0.4736 & 21866565 \\
\hline
\multicolumn{12}{c}{\textbf{EMRI Parameters}} \\
\hline
$M$ & $\mu$ & $a$ & $p_0$ & $e_0$ & $d_L$ & $q_S$ & $\phi_S$ & $q_K$ & $\phi_K$ & $\Phi_{\varphi 0}$ & $\Phi_{r0}$ \\
1099500 & 9.7500 & 0.7950 & 6.5000 & 0.4500 & 0.7500 & 0.7750 & 1.9500 & 1.5500 & 0.9500 & 1.9500 & 0.2750 \\
\hline
\end{tabular*}
\end{table*}

\begin{table*}[htbp]
\centering
\caption{True parameters used for the two EMRI injections
(SNR $=80.51$ and $70.95$; local sequential fit, without noise,
$T_{\rm obs}=1$\,yr).}
\label{tab:true_params_2_EMRI_noise}
\renewcommand{\arraystretch}{1.15}
\begin{tabular*}{\textwidth}{@{\extracolsep{\fill}}lllllllllllll@{}}
\hline
\multicolumn{13}{c}{\textbf{EMRI Parameters}} \\
\hline
Source & $M$ & $\mu$ & $a$ & $p_0$ & $e_0$ & $d_L$ (Gpc) & $q_S$ & $\phi_S$ & $q_K$ & $\phi_K$ & $\Phi_{\varphi 0}$ & $\Phi_{r0}$ \\
\hline
EMRI 1 & 1009500 & 10.2500 & 0.7500 & 7.0000 & 0.3500 & 0.7500 & 0.7750 & 1.9500 & 1.5500 & 0.9500 & 1.9500 & 0.2750 \\
EMRI 2 & 1010500 & 10.7500 & 0.8500 & 8.0000 & 0.4500 & 0.8500 & 0.8250 & 2.0500 & 1.6500 & 1.0500 & 2.0500 & 0.3250 \\
\hline
\end{tabular*}
\end{table*}

\begin{table*}[htbp]
\centering
\caption{Injected parameters for the signal comprising two MBHB mergers and one EMRI.}
\label{tab:MBH_2_EMRI_1_without_noise}
\renewcommand{\arraystretch}{1.15}

\begin{tabular*}{\textwidth}{@{\extracolsep{\fill}}lllllllllllll@{}}
\hline
\multicolumn{13}{c}{\textbf{MBHB Parameters}} \\
\hline
Source & $M_T\,[M_\odot]$ & $q$ & $a_1$ & $a_2$ & $\iota$
& $D_L$ (Gpc) & $\phi_{\rm ref}$ & $\lambda$ & $\beta$
& $\psi$ & $t_{\rm ref}$ (s) \\
\hline
MBHB 1 &
$1.926{\times}10^6$ & 0.581 & 0.647 & 0.739 & 0.547 &
37.5 & 0.700 & 0.443 & 0.803 & 0.474 &
$8.666{\times}10^5$ \\

MBHB 2 &
$1.946{\times}10^6$ & 0.584 & 0.847 & 0.939 & 1.547 &
42.5 & 1.700 & 0.543 & 1.003 & 0.574 &
$2.087{\times}10^7$ \\
\hline

\multicolumn{13}{c}{\textbf{EMRI Parameters}} \\
\hline
Source & $M\,[M_\odot]$ & $\mu\,[M_\odot]$ & $a$ & $p_0$ & $e_0$
& $d_L$ (Gpc) & $q_S$ & $\phi_S$ & $q_K$ & $\phi_K$
& $\Phi_{\varphi0}$ & $\Phi_{r0}$ \\
\hline
EMRI 1 &
$1.100{\times}10^6$ & 9.75 & 0.70 & 7.5 & 0.25 &
0.75 & 0.75 & 1.95 & 1.55 & 0.95 &
1.95 & 0.275 \\
\hline
\end{tabular*}
\end{table*}

\begin{table*}[htbp]
\centering
\caption{Injected parameters for the signal comprising two time-aligned MBHB coalescence and one EMRI.}
\label{tab:MBH_2_EMRI_1_overlapping_MBHs}
\renewcommand{\arraystretch}{1.15}

\begin{tabular*}{\textwidth}{@{\extracolsep{\fill}}llllllllllll@{}}
\hline
\multicolumn{12}{c}{\textbf{MBHB Parameters}} \\
\hline
Source & $M_T\,[M_\odot]$ & $q$ & $a_1$ & $a_2$ & $\iota$
& $D_L$ (Gpc) & $\phi_{\rm ref}$ & $\lambda$ & $\beta$
& $\psi$ & $t_{\rm ref}$ (s) \\
\hline
MBHB 1 &
$1.925762{\times}10^6$ & 0.5807 & 0.6474 & 0.7388 & 0.5472 &
37.5 & 0.700 & 0.4425 & 0.8030 & 0.4736 &
$1.086562{\times}10^7$ \\

MBHB 2 &
$1.945762{\times}10^6$ & 0.5841 & 0.8474 & 0.9388 & 1.5472 &
42.5 & 1.700 & 0.5425 & 1.0030 & 0.5736 &
$1.086762{\times}10^7$ \\
\hline
\end{tabular*}

\vspace{0.8em}

\begin{tabular*}{\textwidth}{@{\extracolsep{\fill}}llllllllllllll@{}}
\hline
\multicolumn{13}{c}{\textbf{EMRI Parameters}} \\
\hline
Source & $M\,[M_\odot]$ & $\mu\,[M_\odot]$ & $a$ & $p_0$ & $e_0$
& $d_L$ (Gpc) & $q_S$ & $\phi_S$ & $q_K$ & $\phi_K$
& $\Phi_{\varphi0}$ & $\Phi_{r0}$ \\
\hline
EMRI 1 &
$1.0995{\times}10^6$ & 9.75 & 0.70 & 8.0 & 0.25 &
0.75 & 0.75 & 1.95 & 1.55 & 0.95 &
1.95 & 0.275 \\
\hline
\end{tabular*}
\end{table*}

\begin{table*}[htbp]
\centering
\caption{True parameters used for the two overlapping EMRI injections; noiseless, $T_{\rm obs}=1$\,yr.  The
SNRs are $79.41$ and $69.4$ respectively.}
\label{tab:true_params_2_EMRI_overlapping_AET}
\renewcommand{\arraystretch}{1.15}
\begin{tabular*}{\textwidth}{@{\extracolsep{\fill}}lllllllllllll@{}}
\hline
\multicolumn{13}{c}{\textbf{EMRI Parameters}} \\
\hline
Source & $M$ & $\mu$ & $a$ & $p_0$ & $e_0$ & $d_L$ (Gpc) & $q_S$ & $\phi_S$ & $q_K$ & $\phi_K$ & $\Phi_{\varphi0}$ & $\Phi_{r0}$ \\
\hline
EMRI 1 & 1009500 & 10.2500 & 0.7500 & 7.0000 & 0.3500 & 0.7500 & 0.7750 & 1.9500 & 1.5500 & 0.9500 & 1.9500 & 0.2750 \\
EMRI 2 & 1010500 & 10.7500 & 0.8500 & 6.7350 & 0.4500 & 0.8500 & 0.8250 & 2.0500 & 1.6500 & 1.0500 & 2.0500 & 0.3250 \\
\hline
\end{tabular*}
\end{table*}

\begin{table*}[htbp]
\centering
\caption{True parameters for the $(4,3)$ missing-mode injections.}
\label{tab:true_params_missing_43_compact}
\renewcommand{\arraystretch}{1.15}
\begin{tabular*}{\textwidth}{@{\extracolsep{\fill}}lllllllllllll@{}}
\hline
\multicolumn{13}{c}{\textbf{MBHB Parameters}} \\
\hline
Case & $M_T$ & $q$ & $a_1$ & $a_2$ & $\iota$ & $D_L$ & $\phi_{\rm ref}$ & $\lambda$ & $\beta$ & $\psi$ & $t_{\rm ref}$ \\
\hline
S1 & 2000000 & 0.2500 & 0.9000 & 0.9000 & 1.3090 & 40.0000 & 1.7000 & 0.5425 & 1.0030 & 0.5736 & 10867620 \\
S2 & 1925762 & 0.5807 & 0.6474 & 0.7388 & 0.5472 & 37.5000 & 0.7000 & 0.4425 & 0.8030 & 0.4736 & 10866565 \\
\hline
\multicolumn{13}{c}{\textbf{EMRI Parameters}} \\
\hline
Case & $M$ & $\mu$ & $a$ & $p_0$ & $e_0$ & $d_L$ & $q_S$ & $\phi_S$ & $q_K$ & $\phi_K$ & $\Phi_{\varphi0}$ & $\Phi_{r0}$ \\
\hline
S1 & 1000000 & 10.0000 & 0.7000 & 8.0000 & 0.2500 & 0.8000 & 0.7500 & 1.9500 & 1.5500 & 0.9500 & 1.9500 & 0.4000 \\
S2 & 1099500 & 9.7500 & 0.7950 & 6.5000 & 0.3500 & 0.7500 & 0.7750 & 1.9500 & 1.5500 & 0.9500 & 1.9500 & 0.2750 \\
\hline
\end{tabular*}
\end{table*}

\begin{table*}[htbp]
\centering
\caption{True parameters for the two missing-mode injections in which both the
$(4,4)$ and $(4,3)$ harmonics are dropped from the MBHB inference template.
Both runs are noiseless, $T_{\rm obs}=1$\,yr.  The two configurations differ only in the MBHB
total mass and luminosity distance; the EMRI injection is identical.}
\label{tab:true_params_missing_44_43}
\renewcommand{\arraystretch}{1.15}
\begin{tabular*}{\textwidth}{@{\extracolsep{\fill}}llllllllllllll@{}}
\hline
\multicolumn{13}{c}{\textbf{MBHB Parameters}} \\
\hline
Case & $M_T$ & $q$ & $a_1$ & $a_2$ & $\iota$ & $D_L$ (Gpc) & $\phi_{\rm ref}$ & $\lambda$ & $\beta$ & $\psi$ & $t_{\rm ref}$ \\
\hline
M1 & $2\times10^{6}$ & 0.2500 & 0.9000 & 0.9000 & 1.3090 & 40.0000 & 1.7000 & 0.5425 & 1.0030 & 0.5736 & 10867620 \\
M2 & $1\times10^{7}$ & 0.2500 & 0.9000 & 0.9000 & 1.3090 & 45.0000 & 1.7000 & 0.5425 & 1.0030 & 0.5736 & 10867620 \\
\hline
\multicolumn{13}{c}{\textbf{EMRI Parameters}} \\
\hline
Case & $M$ & $\mu$ & $a$ & $p_0$ & $e_0$ & $d_L$ (Gpc) & $q_S$ & $\phi_S$ & $q_K$ & $\phi_K$ & $\Phi_{\varphi0}$ & $\Phi_{r0}$ \\
\hline
M1, M2 & 1000000 & 10.0000 & 0.7000 & 8.0000 & 0.2500 & 0.8000 & 0.7500 & 1.9500 & 1.5500 & 0.9500 & 1.9500 & 0.4000 \\
\hline
\end{tabular*}
\end{table*}

\begin{table*}[htbp]
\centering
\caption{True parameters used for the three MBHB and three EMRI injections in
the joint MBHB / sequential EMRI configuration.  The three MBHBs are fitted jointly, after which the three EMRIs are fitted sequentially on the residual.}
\label{tab:true_params_3_MBH_3_EMRI}
\renewcommand{\arraystretch}{1.15}
\begin{tabular*}{\textwidth}{@{\extracolsep{\fill}}lllllllllllll@{}}
\hline
\multicolumn{13}{c}{\textbf{MBHB Parameters}} \\
\hline
Source & $M_T$ & $q$ & $a_1$ & $a_2$ & $\iota$ & $D_L$ (Gpc) & $\phi_{\rm ref}$ & $\lambda$ & $\beta$ & $\psi$ & $t_{\rm ref}$ \\
\hline
MBHB 1 & 1980762 & 0.6882 & 0.2974 & 0.3888 & 0.0472 & 39.4025 & 0.7000 & 0.2425 & 0.6530 & 0.2736 & 10249200 \\
MBHB 2 & 2035762 & 0.6642 & 0.5474 & 0.6388 & 1.0472 & 35.9025 & 1.2000 & 0.4925 & 0.9030 & 0.5236 & 15768000 \\
MBHB 3 & 2090762 & 0.6420 & 0.7974 & 0.8888 & 2.0472 & 32.4025 & 1.7000 & 0.7425 & 1.1530 & 0.7736 & 21286800 \\
\hline
\multicolumn{13}{c}{\textbf{EMRI Parameters}} \\
\hline
Source & $M$ & $\mu$ & $a$ & $p_0$ & $e_0$ & $d_L$ (Gpc) & $q_S$ & $\phi_S$ & $q_K$ & $\phi_K$ & $\Phi_{\varphi0}$ & $\Phi_{r0}$ \\
\hline
EMRI 1 & 1010000 & 10.5000 & 0.6000 & 8.0000 & 0.2500 & 1.0000 & 0.8000 & 2.0000 & 1.6000 & 1.0000 & 2.0000 & 0.3000 \\
EMRI 2 & 985000 & 10.0000 & 0.3500 & 7.5000 & 0.1000 & 0.7500 & 0.3000 & 1.5000 & 1.1000 & 0.5000 & 1.5000 & $-0.2000$ \\
EMRI 3 & 1035000 & 11.0000 & 0.8500 & 8.5000 & 0.4000 & 1.2500 & 1.3000 & 2.5000 & 2.1000 & 1.5000 & 2.5000 & 0.8000 \\
\hline
\end{tabular*}
\end{table*}

\clearpage
\section{Posteriors}
    \label{appendix_B}
    
\begin{figure}[H]
    \centering
    \includegraphics[scale = 0.30]{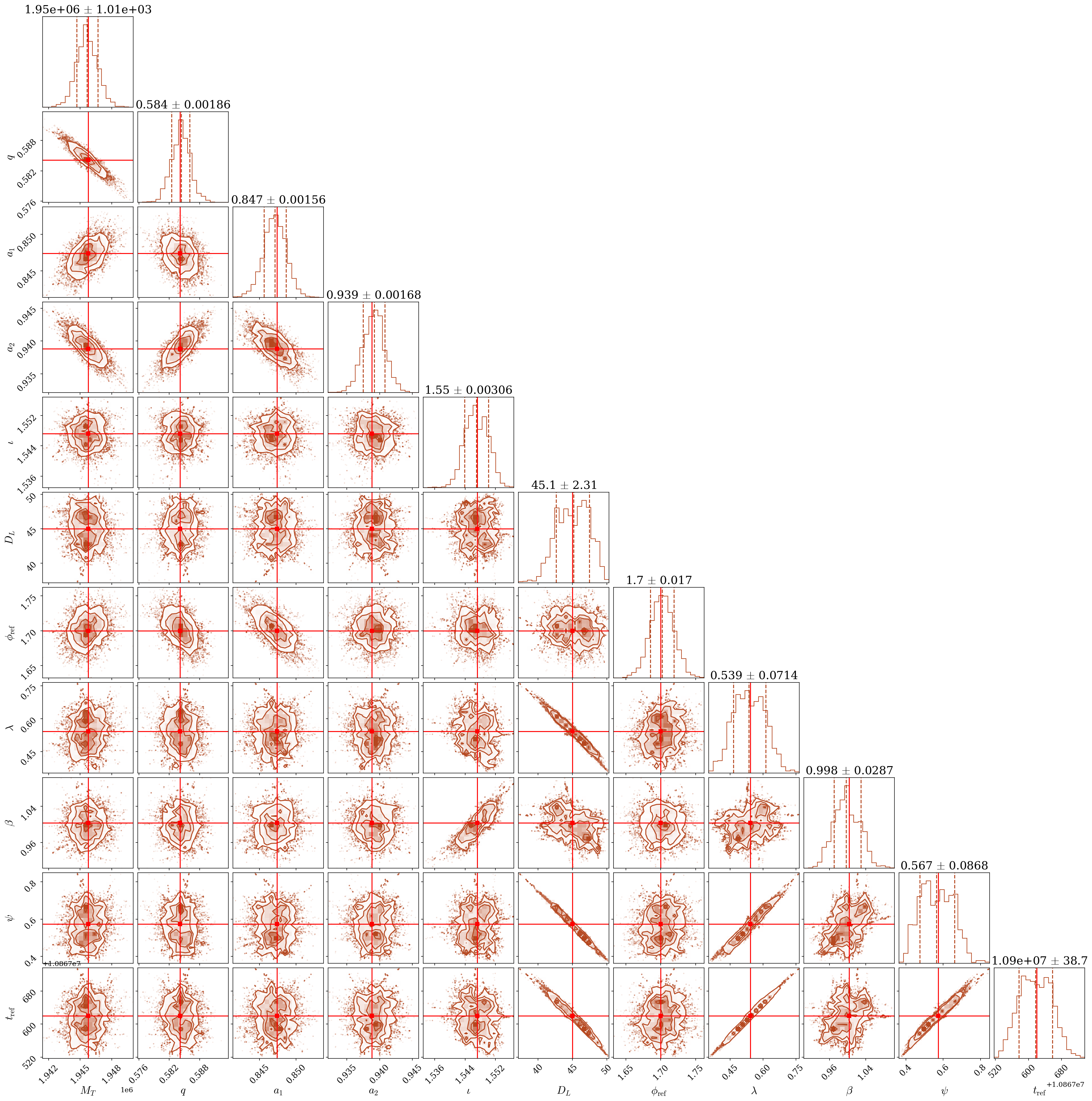}
    \caption{Posterior distributions of the MBHB parameters inferred in the presence of a loud EMRI signal in the \textit{LoudEMRI} configuration. Some one-dimensional marginalized posteriors exhibit noticeable departures from Gaussianity.}
    \label{fig:MBHB_characterization_loud_EMRI}
\end{figure}
\begin{figure}[H]
    \centering
    \includegraphics[scale = 0.27]{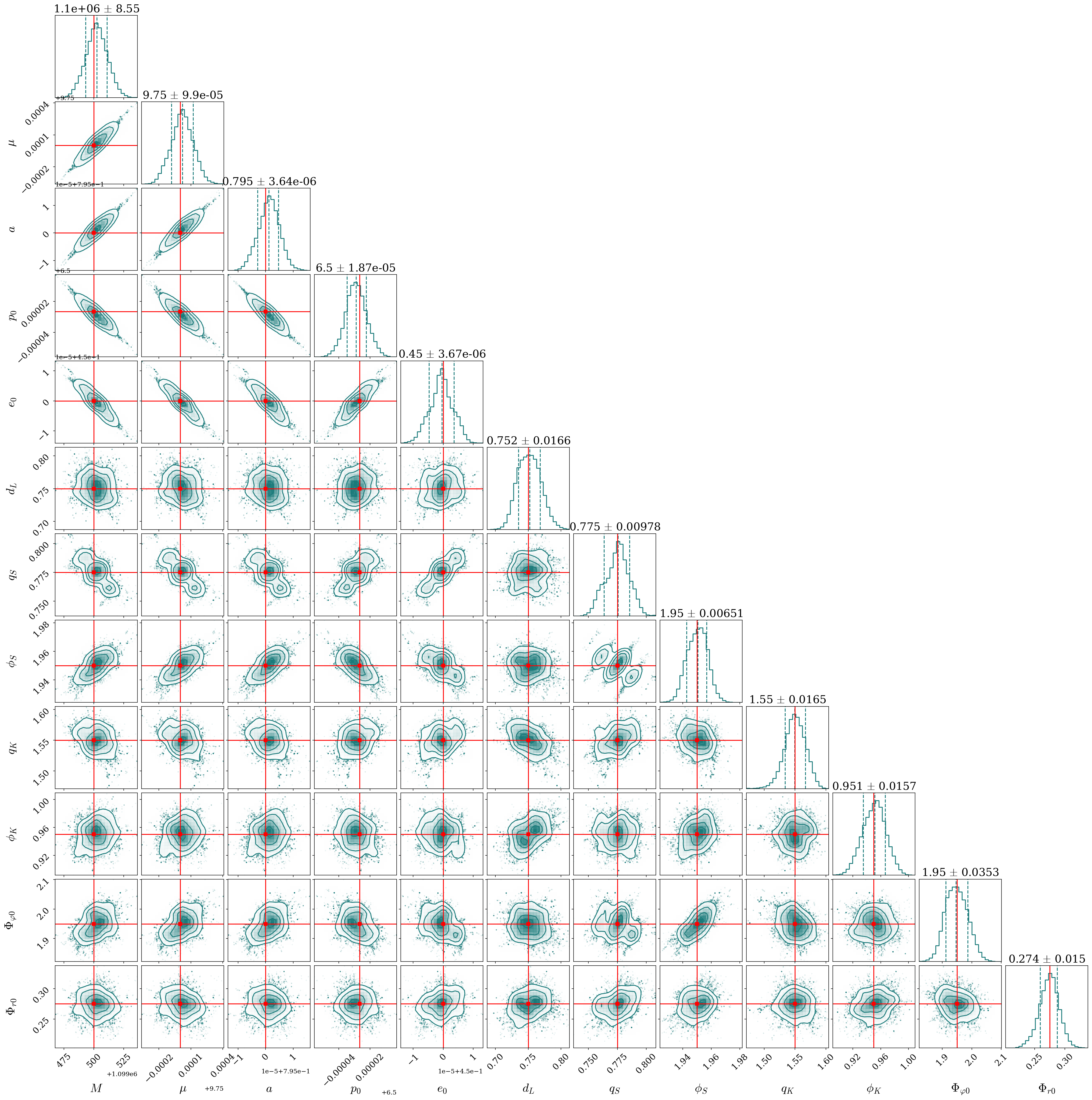}
    \caption{Posterior distributions of the EMRI parameters inferred in the \textit{Time-Aligned Plunge} configuration, where the EMRI plunge approximately coincides with the MBHB merger. Some one-dimensional marginalized posteriors exhibit noticeable departures from Gaussianity.}
    \label{fig:corner_plot_1emri_1mbhb_overlapping}
\end{figure}
\clearpage            
\twocolumngrid        
\bibliographystyle{apsrev4-2}
\bibliography{reference_new}

\end{document}